\documentclass[prd,twocolumn,preprintnumbers,showpacs,nofootinbib,superscriptaddress,notitlepage,floatfix]{revtex4-1}
\usepackage{hyperref}
\usepackage{graphicx}
\usepackage{amsmath}
\usepackage{amssymb}
\usepackage{amsfonts}
\usepackage{xcolor}
\usepackage{multirow}

\DeclareRobustCommand{\Eq}[1]{Eq.~\eqref{eq:#1}}
\DeclareRobustCommand{\eq}[1]{Eq.~\eqref{eq:#1}}

\DeclareRobustCommand{\eqs}[2]{Eqs.~\eqref{eq:#1} and \eqref{eq:#2}}

\DeclareRobustCommand{\fig}[1]{Fig.~\ref{fig:#1}}

\DeclareRobustCommand{\app}[1]{App.~\ref{app:#1}}
\DeclareRobustCommand{\sec}[1]{Sec.~\ref{sec:#1}}

\DeclareRobustCommand{\tbl}[1]{Table~\ref{tbl:#1}}
\DeclareRobustCommand{\refcite}[1]{Ref.~\cite{#1}}

\DeclareRobustCommand{\eq}[1]{Eq.~(\ref{eq:#1})}
\DeclareRobustCommand{\eqs}[2]{Eqs.~(\ref{eq:#1}) and (\ref{eq:#2})}

\newcommand{\MS}{{\overline{\mathrm{MS}}}}

\newcommand{\eps}{\epsilon}

\newcommand{\nn}{\nonumber}

\newcommand\bets{\begin{table*}}
\newcommand\eets[1]{\label{tb:#1}\end{table*}}

\begin{document}

%%%%%%%%%%%%%%%%%%%%%%%%%%%%%%%%%%%%%%%%%%%%%%%%%%%%%%%%%%%%%%%%%%%%%%
\title{Origin and Resummation of Threshold Logarithms in the\\ Lattice QCD Calculations of PDFs}
%%%%%%%%%%%%%%%%%%%%%%%%%%%%%%%%%%%%%%%%%%%%%%%%%%%%%%%%%%%%%%%%%%%%%%

%%%%%%%%%%%%%%%%%%%%%%%%%%%%%%%%%%%%%%%%%%%%%%%%%%%%%%%%%%%%%%%%%%%%%%

\author{Xiang Gao}
\email{xgao@bnl.gov}
\affiliation{Physics Department, Brookhaven National Laboratory, Bldg. 510A, Upton, NY 11973, USA}
\affiliation{Physics Department, Tsinghua University, Beijing 100084, China}

\author{Kyle Lee}
\email{kylelee@lbl.gov}
\affiliation{Nuclear Science Division, Lawrence Berkeley National Laboratory, Berkeley, California 94720, USA}
\affiliation{Physics Department, University of California, Berkeley, CA 94720, USA}

\author{Swagato Mukherjee}
\affiliation{Physics Department, Brookhaven National Laboratory, Bldg. 510A, Upton, NY 11973, USA}

\author{Charles Shugert}
\affiliation{Physics Department, Brookhaven National Laboratory, Bldg. 510A, Upton, NY 11973, USA}
\affiliation{Department of Physics and Astronomy, Stony Brook University, Stony Brook, NY 11794, USA}

\author{Yong Zhao}
\email{yzhao@bnl.gov}
\affiliation{Physics Department, Brookhaven National Laboratory, Bldg. 510A, Upton, NY 11973, USA}
\affiliation{Physics Division, Argonne National Laboratory, Lemont, IL 60439, USA}

\begin{abstract}

Many present lattice QCD approaches to calculate the parton distribution functions (PDFs) rely on a factorization formula or effective theory expansion of certain Euclidean matrix elements in boosted hadron states. In the quasi- and pseudo-PDF methods, the matching coefficient in the factorization or expansion formula includes large logarithms near the threshold, which arise from the subtle interplay of collinear and soft divergences of an underlying 3D momentum distribution. We use the standard prescription to resum such logarithms in the Mellin-moment space at next-to-leading logarithmic accuracy, which also accounts for the DGLAP evolution, and we show that it can suppress the PDF at large $x$. Unlike the deep inelastic scattering and Drell-Yan cross sections, the resummation formula is away from the Landau pole. We then apply our formulation to reanalyze the recent lattice results for the pion valence PDF, and find that within the current data sensitivity, the effect of threshold resummation is marginal for the accessible moments and the PDF at large $x$.

\end{abstract}

\maketitle

\section{Introduction}

Parton distribution functions (PDFs) are among the most basic quantities that describe the inner structure of the proton in high-energy experiments such as the deep inelastic scattering (DIS) and Drell-Yan (DY) processes. Being intrinsic to the hadrons, which is dominated by the strong interaction, the PDFs are nonperturbative functions and can only be determined from experiments or first-principles calculations in Quantum Chromodynacmis (QCD).

Determination of the PDFs from experimental measurements is based on QCD factorization theorems~\cite{Collins:1989gx}, where a cross section can be factorized as a convolution of a hard coefficient function, which can be calculated analytically in perturbation theory, and a PDF. For instance, the factorization of the DIS cross section of a lepton scattering off of a hadron of momentum $P$, with the exchange of a boson of momentum $q$, can be schematically written as
\begin{align}
\label{eq:DISfact}
\frac{1}{\sigma_0}\frac{d\sigma_{\rm DIS}}{d\eta dQ^2} =\eta \int_\eta^1 \frac{dz}{z}\ C\left(z,Q,\mu\right)q\left(\frac{\eta}{z},\mu\right)\,,
\end{align}
where $Q^2=-q^2$, $\eta = Q^2/(2P\cdot q)$ is the Bjorken variable and $\mu$ is the factorization scale. Here $q(x,\mu)$ is a PDF at the momentum fraction $x$, and $C$ is the hard coefficient function.
Using similar factorization formulas for different processes to fit to the enormous amount of cross section data accumulated from the worldwide high-energy physics programs over the past few decades, the PDFs have been extracted with remarkable precision~\cite{Martin:2009iq,Harland-Lang:2014zoa,Buckley:2014ana,Dulat:2015mca,Alekhin:2017kpj,Ball:2017nwa}. Nevertheless, due to the limitations of experiments, there are still large uncertainties for the PDFs of the sea quarks, gluon, spin-dependent partons, and at small and large $x$ regions.

Among them, the extraction of the PDF at large $x$ is complicated by the soft divergences in the hard coefficient functions as their partonic variables approach threshold, for instance, $C(z,Q,\mu)$ of \eq{DISfact} as $z$ approaches one. Such divergence gives rise to threshold logarithms of $(1-z)$, which necessitates all-order resummation~\cite{Sterman:1986aj,Catani:1989ne,Becher:2006mr}. The global analysis of the PDF is then affected by whether one uses the fixed-order or the resummed version of the coefficient function $C$~\cite{Corcella:2005us,Aicher:2010cb,Bonvini:2015ira,Westmark:2017uig}. As can be seen from Eq.\ \eqref{eq:DISfact}, since the PDF is in general convoluted with a hard coefficient function, the extraction of the PDF at even moderately large $x$ may be affected by threshold resummation.
For example, the NNPDF collaboration used DIS, DY, and inclusive top-quark pair production processes simultaneously to study the impact of threshold resummation in PDF extractions~\cite{Bonvini:2015ira}, and concluded that it suppresses the PDF in the large-$x$ region around $x \gtrsim 0.1$, while at intermediate values of $x$, $0.01 \lesssim x \lesssim 0.1$, the quark PDFs are somewhat enhanced\footnote{Note that although the central value of the PDF departs from that obtained without resummation, they
% find that the resummed fits still have large uncertainties which overlaps with the fixed order fits
still agree within large uncertainties due to the lack of experimental data in the relevant kinematic region.}. The suppression of the extracted PDFs at large $x$ can be intuitively understood as compensation for the enhancement of resummed coefficient functions near threshold.

Among the many PDFs extracted from experiments, those of the pions are of special interests as they are the pseudo-Nambu-Goldstone bosons of QCD and are the simplest hadrons for studying strong dynamics. For example, the behavior of the pion valence PDFs at large $x$, which is characterized by a power law behavior $(1-x)^\beta$, has been under debate among various approaches. As $x\to1$, the pion momentum is mostly carried by the active valence quark, so the spectator quark is at low energy and interacts strongly with other soft particles. In the asymptotic limit, the Brodsky-Farrar quark counting rules~\cite{Brodsky:1973kr} predicted that $\beta\sim 2$, but the analysis of the E-0615 Fermilab data suggested $\beta\sim$ 1~\cite{Conway:1989fs}. With the inclusion of threshold resummation, the Fermilab data was reanalyzed by the authors of Ref.~\cite{Aicher:2010cb} (ASV), and they found $\beta\sim$ 2. Moreover, the recent global Monte Carlo analysis from the JAM collaboration\ \cite{Barry:2018ort}, as well as the analysis done using the {\UrlFont xFitter} program\ \cite{Novikov:2020snp}, suggest that $\beta\sim 1$, although neither of them included threshold resummation.

Apart from the determination of the PDFs using experimental data, there have also been long-standing efforts to calculate the PDFs from the first-principles method of lattice QCD. Due to their explicit dependence on the real time, the PDFs cannot be directly simulated on the Euclidean lattice that uses imaginary time. Instead, their Mellin moments, which are matrix elements of local gauge-invariant operators, can be calculated on the lattice. However, these calculations are only viable for the lowest few moments due to power divergent operator mixings and worsening of the signal-to-noise ratio for the higher moments (for a review, see Ref.~\cite{Lin:2017snn}).

In recent years, several approaches have been proposed to calculate either the higher Mellin moments or $x$-dependence of the PDF. They include the calculation of the hadronic tensor~\cite{Liu:1993cv,Liu:1998um,Liu:1999ak,Liu:2017lpe}, restoration of rotational symmetry to access higher moments~\cite{Davoudi:2012ya}, operator product expansion (OPE) of a Compton amplitude in the unphysical region~\cite{Detmold:2005gg,Chambers:2017dov}, the coordinate-space OPE or short-distance expansion of the current-current correlator~\cite{Braun:2007wv,Ma:2017pxb}, the quasi-PDF (qPDF) in large-momentum effective theory (LaMET)~\cite{Ji:2013dva,Ji:2014gla,Ji:2020ect}, and the pseudo-PDF (pPDF) approach~\cite{Radyushkin:2017cyf,Orginos:2017kos}.
Except for the first two approaches, all the other ones are based on the factorization of a Euclidean observable into the PDF (or its moments) and a perturbative matching coefficient, up to power-suppressed corrections. Among them, the LaMET approach has attracted wide interest and led to much progress towards systematic calculations of the $x$-dependence of the PDF over the past few years. Within this approach, the PDF can be obtained from the qPDF, which is the Fourier transform of a spatial correlator in a boosted hadron state, through an effective theory expansion in powers of large hadron momentum~\cite{Ji:2020ect}. The pPDF approach is closely related to LaMET as it uses an equivalent short-distance expansion of the same spatial correlator in the coordinate space~\cite{Ji:2017rah,Radyushkin:2017ffo,Izubuchi:2018srq}. The pPDF in this approach is a different Fourier transform of the same spatial corrrelator.

In both the large-momentum expansion for the qPDF and the short-distance expansion for the pPDF (or spatial correlator), one suffers from power corrections at the end-point regions as $x\to0$ and $x\to1$~\cite{Izubuchi:2018srq,Braun:2018brg,Liu:2020rqi,Ji:2020brr}. Apart from the power corrections, the matching coefficient in the leading-power contribution includes large logarithms in the threshold region, which will also impact the determination of the large-$x$ exponent for the PDF. Similar large logarithms are also observed in the matching coefficient for the short-distance expansion of the current-current correlator~\cite{Sufian:2020vzb}, which, however, is not the focus of this work. The forms of such logarithms are very similar to the threshold logarithms in the DIS and DY cross section formulas, but their origins are not yet understood, especially for the pPDF or spatial correlator, since there is no clear correspondence to a soft-momentum limit in its variables. As will be shown below, both the qPDF and the pPDF are the special limits of a 3D momentum distribution defined with a straight gauge link. This 3D momentum distribution then has a dependence on the variable $x$ with clear physical interpretation as momentum fraction, and includes threshold logarithms of $(1-x)$ in the limit of $x \to 1$. The relation between this 3D distribution and the qPDF or the pPDF then elucidates the origin of the large logarithms found in the matching coefficients mentioned above. Especially, the threhsold logarithm in the matching coefficient for pPDF matches the factorization scale to the ultraviolet (UV) fixed point in the $x\to1$ limit, which is opposite to the DIS and DY cases where the scale being matched to becomes infrared (IR) in this limit.
After identifying the origin of these threshold logarithms, we will then resum them at next-to-leading logarithmic (NLL) accuracy by using the form motivated by the standard threshold resummation formulas in the literature~\cite{Catani:1996yz}, and also confirm our resummation formula with the next-to-next-to-leading order (NNLO) matching coefficients~\cite{Li:2020xml}. Unlike the DIS and DY cases, the resummed matching coefficient for the pPDF or spatial correlator is free from the Landau pole as it matches to the UV fixed point.

Since the large-$x$ region of the PDF is sensitive to long-range correlations, which are usually beyond the reach of contemporary lattice data, the predictive power of lattice QCD is limited. Although two existing lattice calculations have tried to fit the large-$x$ exponent of the pion valence PDF~\cite{Sufian:2020vzb,Gao:2020ito}, neither of them has quantified the power corrections or threshold resummation effects, and their results were inconclusive. In this work, we investigate the effects of threshold resummation on the fitted Mellin moments and the large-$x$ exponent of the pion valence PDF for the first time. Despite the fact that current lattice data are only sensitive to lower moments and PDF within a moderate range of $x$, this investigation is aimed to deepen our understanding of the systematics at large $x$ and to provide the framework for future analyses of data that are sensitive to this region.

The rest of this work is presented as follows: In \sec{origin}, we show that threshold logarithms in the qPDF and pPDF (or spatial correlator) originate from the leading divergences in the 3D momentum distribution as $x\to1$. As a byproduct, we find that the qPDF and pPDF (or spatial correlator) are special limits of this particular 3D distribution, revealing a new relation between the two in addition to Fourier transforms. Then in \sec{nllTR}, we resum the threshold logarithms at NLL accuracy with next-to-leading order (NLO) matching, and estimate its impact on the large-$x$ exponent of the PDF. In \sec{num}, we use the resummed formula to fit the second and fourth moments, as well as to fit the pion valence PDF directly with the parametrization $\sim x^{\alpha}(1-x)^\beta$, and show that the effect of threshold resummation on the moments or value of $\beta$ is negligible compared to other uncertainties in the current lattice data. Finally, we conclude in \sec{sum}.

\section{Origin of threshold logarithms}
\label{sec:origin}

In QCD, the PDF $q(x,\mu)$ is defined from the Fourier transform of the light-front correlator $h_{\gamma^+}(\lambda,\mu)$,
\begin{align}
    q(x,\mu) &= \int_{-\infty}^\infty \frac{d\lambda}{2\pi} e^{-i\lambda x} h_{\gamma^+}(\lambda,\mu)\,,\\
    h_{\gamma^+}(\lambda,\mu) &= \frac{1}{2P^+} \langle P| \bar{\psi}(\xi^-)W(\xi^-,0) \gamma^+ \psi(0)|P\rangle\,,
\end{align}
where $P^\mu=(P^t,0,0,P^z)$ and $\lambda=P^+\xi^-$, with light-cone coordinates $\xi^\pm = (t\pm z)/\sqrt{2}$. Here $h_{\gamma^+}(\lambda,\mu)$ is renormalized in the $\MS$ scheme at renormalization scale $\mu$. The gauge link that points from $x_1$ to $x_2$ is defined as a path integration
\begin{align}
    &W(x_2,x_1) = \\
    &P\exp\left(-ig \int_0^1 ds\ (x_2-x_1)_\mu  A^\mu \big(x_1+(x_2-x_1)s\big)\right)\,,\nn
\end{align}
where $A^\mu=A^\mu_a t^a$ is in the fundamental representation of $SU(3)$.

Within the LaMET approach~\cite{Ji:2013dva,Ji:2014gla,Ji:2020ect}, $q(x,\mu)$ can be extracted from the qPDF that is defined from a spatial quark correlator in a boosted hadron state,
\begin{align}\label{eq:qpdf}
\tilde{q}(x,P^z,\mu) &\equiv \int_{-\infty}^\infty \frac{d\lambda}{2\pi} e^{i\lambda x} \tilde h_{\Gamma}(\lambda,P^z,\mu)\,,\\
\label{eq:spatial}
\tilde h_{\Gamma}(\lambda,P^z,\mu) &\equiv \frac{1}{N_\Gamma} \langle P| \bar{\psi}(b)W(b,0) \Gamma \psi(0)|P\rangle\,,
\end{align}
where $b^\mu=(0,0,0,z)$, $\lambda= - b \cdot P = z P^z$, and $\Gamma=\gamma^z$ or $\gamma^t$ with $N_{\gamma^z}=2P^z$ and $N_{\gamma^t}=2P^t$. When $P^z \gg \Lambda_{\rm QCD}$, the qPDF (for the non-singlet case) is related to the PDF through the factorization formula~\cite{Xiong:2013bka,Ma:2014jla,Izubuchi:2018srq}
\begin{align} \label{eq:factorization}
\tilde{q}(x, P^z, \mu) & =  \int_{-1}^{1} \frac{dy}{|y|} \ C\!\left(\frac{x}{y}, \frac{\mu}{yP^z}\right) q(y,\mu)+\ldots \,,
\end{align}
where the region $-1<y<0$ corresponds to the antiquark distribution through $q(y)=-\bar{q}(-y)$. Here $C$ is a perturbative matching coefficient that matches the UV difference between the qPDF and PDF, and depends on the logarithm of $\mu/(yP^z)$. The ``$\ldots$'' denotes target-mass~\cite{Chen:2016utp} and higher-twist corrections that are generally of ${\cal O}\left( \Lambda_{\rm QCD}^2/(xP^z)^2, \Lambda_{\rm QCD}^2/((1-x)P^z)^2\right)$~\cite{Ji:2020brr}.

The spatial correlator $\tilde h_{\Gamma}(\lambda, P^z,\mu)$ can also be written as a function of Lorentz scalars $\lambda$ and $b^2 = -z^2 $~\cite{Radyushkin:2017cyf}, i.e., $\tilde h_{\Gamma}(\lambda, -b^2\mu^2)$. At short distance, $\tilde h_{\Gamma}(\lambda, -b^2\mu^2= z^2\mu^2)$ can be matched to the light-cone correlator $\tilde h_{\gamma^+}(\lambda,\mu)$ through the factorization formula~\cite{Ji:2017rah,Radyushkin:2017lvu,Izubuchi:2018srq}
\begin{align}
\label{eq:io-Q-fact}
\tilde{h}_{\Gamma}(\lambda, z^2\mu^2)
&=\!\int_{-1}^1 dw\ \mathcal{C}(w, z^2\mu^2)\ h_{\gamma^+}(w\lambda,\mu) \!+ \ldots \,,
\end{align}
where ${\cal C}$ is the matching coefficient, and ``$\ldots$'' denote the target-mass and higher-twist corrections of ${\cal O}(z^2\Lambda^2_{\rm QCD})$.

The pPDF ${\cal P}(x,z^2\mu^2)$ is related to the above spatial correlator through Fourier transform with $z^2$ fixed,
\begin{align}
    {\cal P}(x,z^2\mu^2) \equiv \int_{-\infty}^\infty \frac{d\lambda}{2\pi} e^{i\lambda x} \tilde h_{\Gamma}(\lambda, -b^2\mu^2= z^2\mu^2)\,,
\end{align}
which also has a short-distance expansion~\cite{Ji:2017rah,Radyushkin:2017lvu,Izubuchi:2018srq}
\begin{align}\label{eq:pseudofac}
    \mathcal{P}(x,z^2\mu^2) &= \int_{|x|}^1 \frac{dy}{|y|}\ {\cal C} \left( \frac{x}{y},z^2\mu^2 \right) q(y,\mu) \\
    &\quad + \int^{-|x|}_{-1} \frac{dy}{|y|}\ {\cal C} \left( \frac{x}{y},z^2\mu^2 \right) q(y,\mu) + \ldots\nn
\end{align}
with the same matching coefficient ${\cal C}(\omega=x/y,z^2\mu^2)$ as the spatial correlator in \eq{io-Q-fact}.

The $\MS$ matching coefficients ${\cal C}$ and $C$ both have been derived up to two-loop order~\cite{Ji:2017rah,Izubuchi:2018srq,Radyushkin:2017lvu,Chen:2020ody,Li:2020xml}. At one-loop order, the matching coefficients in the threshold limit behave as
\begin{align}
	&\lim_{\xi\to1}C^{(1)}\left(\xi, {\mu\over p^z}\right) \nn\\
	&= {\alpha_sC_F\over 2\pi}\left[ \left({2\ln|1-\xi|\over |1-\xi|}\right)_{\oplus}  - {2\over (1-\xi)_+} \ln{\mu^2\over 4p_z^2}\theta(1-\xi) \right.\nn\\
	&\qquad\qquad\qquad \left. + {3\over 2|1-\xi|_{\oplus}}\right]\,,\label{eq:quasi}\\
&\lim_{w\to1}{\cal C}^{(1)}(w,-b^2\mu^2)\nn\\
&= {\alpha_sC_F\over 2\pi}\left\{\left({2\over 1-w}\right)_{+}\left[-\ln(-b^2\mu^2)-\ln{e^{2\gamma_E}\over 4}-1\right]\right.\nn\\
&\qquad\qquad\left. - \left(4\ln(1-w)\over 1-w\right)_{+}\right\}\!\theta(w)\theta(1-w)
\,,\label{eq:pseudo}
\end{align}
where $C_F=4/3$, and $\xi$ can approach one from both above and below. The plus functions labelled by ``$+$'' are the usual ones defined within the region $[0,1]$, while the definition of those labelled by ``$\oplus$'' is
\begin{align}\label{eq:plus}
	\left({g(x) \over |x|}\right)_{\oplus} &= {g(x) \over |x|} - \delta(x) \int_{-1}^1 dx' {g(x') \over |x'|}
\end{align}
for arbitrary function $g(x)$ with $x\in (-\infty, \infty)$.

As one can see, both \eqs{quasi}{pseudo} include threshold logarithm structures of the form $\ln(1-x)/(1-x)$. In the qPDF case, the matching coefficient $C$ matches the factorization scale $\mu$ to $(1-\xi)^{1/2}p^z$, which becomes IR as $\xi\to1$, just like the DIS and DY cross sections. This can be understood as the probability for a collinear quark to emit a soft gluon with momentum fraction $1-\xi$, which is sensitive to IR physics.
In contrast, the matching coefficient ${\cal C}$ for the spatial correlator and pPDF matches $\mu$ to $(1-\omega)^{-1}z_0^{-1}$ with $z_0=|z|e^{\gamma_E}/2$, which becomes the UV fixed point in the limit of $\omega\to1$. Since the variable $\omega$ does not have a direct physical interpretation as momentum fraction, the physics of this UV fixed point is unclear at first sight.
As we will see later, the resummation of such logarithms will be very different from the DIS and DY cases. Therefore, it is important to understand the physical origin of the threshold logarithms for qPDF and pPDF.

To understand their origins, let us consider a 3D momentum distribution~\cite{Musch:2010ka,Ji:2020ect},
\begin{align}\label{eq:tmd}
	\tilde q(x,\vec{k}_\perp,P^z) &={1\over 2P^0} \int {dz d^2 \vec{b}_\perp\over (2\pi)^3} e^{i\vec{k}_\perp\cdot \vec{b}_\perp + iz (xP^z)}\nonumber \\
	&\quad\times\langle P | \bar{\psi}(b) W(b,0)\gamma^t\psi(0)| P\rangle\,,
\end{align}
where $b^\mu=(0,\vec{b}_\perp, z)$. We suppress the $\mu$-dependence as the following results apply to both bare and renormalized quantities. Here, $\tilde q(x,\vec{k}_\perp,P^z)$ is different from the standard transverse-momentum-dependent (TMD) distribution as it is defined with a straight gauge link, instead of the staple-shaped one in the semi-inclusive DIS or DY processes \cite{Belitsky:2002sm,Boer:2003cm}.

This new straight-gauge-link TMD can also be Fourier transformed to the transverse coordinate space as,
\begin{align}\label{eq:ft}
    \tilde q(x,\vec{b}_\perp,P^z) &= \int d^2\vec{k}_\perp\ e^{-i\vec{k}_\perp\cdot \vec{b}_\perp}\tilde q(x,\vec{k}_\perp,P^z)\,.
\end{align}
Therefore, the 3D momentum distribution can be related to the qPDF $\tilde q(x,P^z)$ through
\begin{align}\label{eq:limits0}
    \int d^2\vec{k}_\perp\ \tilde q(x,\vec{k}_\perp,P^z) &= \tilde q(x, P^z)
\end{align}
or
\begin{align}\label{eq:limits1}
    \lim_{b_\perp\to0}\tilde q(x,\vec{b}_\perp,P^z) &= \tilde q(x, P^z)\,,
\end{align}
which trivially satisfies the definition in \eq{qpdf}. Therefore, the variable $x$ in the qPDF has the physical meaning as a momentum fraction, and the $x\to1$ limit corresponds to $\xi\to 1$ in \eq{quasi}, which stands for the emission of a soft gluon from a collinear quark. Therefore, the physical origin of the threshold logarithms in the qPDF is identical to the usual case.

As for the spatial correlator $\tilde h_{\gamma^t}(\lambda, -b^2)$ in \eq{spatial}, we first note that in the infinite momentum limit,
\begin{align}\label{eq:limits2}
	\lim_{P^z\to\infty }\tilde q(x,\vec{b}_\perp,P^z) &= \int^{\infty}_{-\infty} {d\lambda \over 2\pi}\ e^{i\lambda x}\tilde h_{\gamma^t}(\lambda, -b^2=\vec{b}_\perp^2)\,,
\end{align}
where for any finite $\lambda=zP^z$, the $P^z\to\infty$ limit leads to $z\to0$ so that $b^2\to-\vec{b}_\perp^2$ is completely decoupled from $\lambda$. Moreover, due to Lorentz invaraince, $\tilde h_{\gamma^t}(\lambda, \vec{b}_\perp^2)$ can also be obtained with $b^\mu=(0,b^-,\vec{b}_\perp)$ in light-cone coordinates, which is exactly the correlator that defines the primordial TMD ${\cal F}(x,\vec{k}_\perp^2)$ introduced in Ref.~\cite{Radyushkin:2017cyf},
\begin{align}
    {\cal F}(x,\vec{k}_\perp^2) &\equiv \int^{\infty}_{-\infty} {d\lambda d^2\vec{b}_\perp\over (2\pi)^3}\ e^{i\lambda x + i\vec{k}_\perp\cdot \vec{b}_\perp}\tilde h_{\gamma^t}(\lambda,\vec{b}_\perp^2)\,,
\end{align}
where $x$ is restricted to the physical region $[-1,1]$.
Moreover, the pPDF $\mathcal{P}(x,\vec{b}_\perp^2)$ is related to the primordial TMD through a Fourier transform in the transverse plane~\cite{Radyushkin:2017cyf},
\begin{align}
    \mathcal{P}(x,\vec{b}_\perp^2) &= \int d^2\vec{k}_\perp \ e^{-i\vec{k}_\perp\cdot \vec{b}_\perp}{\cal F}(x,\vec{k}_\perp^2)\,.
\end{align}

Therefore, we can identify $\tilde q(x,\vec{k}_\perp,P^z=\infty)$ as the primordial TMD and $\tilde q(x,\vec{b}_\perp,P^z=\infty)$ as the pPDF, thus the variable $x$ always has a physical interpretation as the longitudinal momentum fraction. Finally, if we set $\vec{b}_\perp^2=z^2$, then $\tilde h_{\gamma^t}(\lambda, \vec{b}_\perp^2)$ coincides with the spatial correlator in \eq{spatial}. Therefore, the variable $\omega$ in the matching coefficient ${\cal C}(\omega,z^2\mu^2)$ in \eqs{io-Q-fact}{pseudofac} indeed corresponds to the longitudinal momentum fraction, which lays the physical ground to explain the origin of the threshold logarithms.\\

Now we use a one-loop calculation of the 3D momentum distribution to explicitly demonstrate the emergence of threshold logarithms of $(1-x)$ in the qPDF and pPDF through relations in Eqs.~(\ref{eq:limits0}--\ref{eq:limits2}). For an on-shell massless external quark state with momentum $p^\mu=(p^z,0,0,p^z)$, the leading collinear and soft divergence of the one-loop 3D momentum distribution comes from the term
\begin{align}\label{eq:sc}
    &\tilde q^{(1)}_{\rm cs}(x,\vec{k}_\perp,p^z)\\
    &= {g^2\mu^{2\eps}C_F\over 2(2\pi)^{d-1}} \int_0^1 ds\ {(1-s)^{2-d}\over \vec{k}_\perp^2}\nn\\
    &\times \left[\frac{k_t^2 (1\!+\!x\!-\!2 s)\!+\!(x-s)^3}{\left( k_t^2+(s-x)^2\right)^{3/2}}\!-\!\frac{k_t^2 (\!1\!+\!x\!-\!2 s)\!+\!(x-1)^3}{\left( k_t^2+(x-1)^2\right)^{3/2}}\right]\,,\nn
\end{align}
where $k_t^2=\vec{k}_\perp^2/p_z^2$, and we work under dimensional regularization with $d=4-2\eps$.

To obtain the qPDF, we integrate over $k_\perp$ and find that
\begin{align}\label{eq:sc1}
    \tilde q^{(1)}_{\rm cs}(x,p^z) &=\int d^{d-2}k_\perp\ \tilde q^{(1)}_{\rm cs}(x,\vec{k}_\perp,p^z)\nn\\
    &\overset{\ x\to 1^-}{\Longrightarrow} - {g^2 C_F\over 8\pi^2}{1\over \epsilon}  {1+(1-x)^{-2\eps}\over 1-x}\big({\mu^2\over p_z^2}\big)^\eps\,,
\end{align}
where the collinear divergence is regulated by $\eps$, and the limit $x\to1$ is approached from below as $x\in(-\infty,\infty)$ for the qPDF~\cite{Xiong:2013bka}. By expanding in $\eps$, we can reproduce the leading threshold logarithm in \eq{quasi}. Here the factor $(1-x)^{-2\eps}$ is crucial to reproduce the correct sign of leading threshold logarithm, which plays a similar role as the phase-space measure in DIS and DY cross sections~\cite{Bolzoni:2007ke}.
In the hard sub-process in DIS and DY, consider a single emission from the outgoing particle, then the two-particle phase-space measure with incoming momentum $\underline{P}$ and outgoing momenta $\underline{Q}$ and $k$ in $d$ dimensions is
\begin{align}
    d\phi_2(\underline{P};\underline{Q},k) &={(\underline{P}^2)^{-\eps}\over 2(4\pi)^{2-2\eps}}  \Big(1-{\underline{Q}^2\over \underline{P}^2}\Big)^{1-2\eps}d\Omega_{d-1}\,,
\end{align}
where $\Omega_{d-1}$ is the solid angle in $(d-1)$ dimensions. For DIS-like emissions, $\underline{P}^2\propto (1-z)$ with $z$ introduced in \eq{DISfact}, and $\underline{Q}^2=0$ corresponds to the emission of a massless particle, i.e., gluon; for DY-like emissions, $(1-{\underline{Q}^2/ \underline{P}^2})\propto (1-z)$, and $\underline{P}^2$ is equal to the threshold value.
The exponent $-\eps$ or $-2\eps$ is responsible for the IR logarithms in the threshold limit.

To obtain the pPDF or primordial TMD, we take the limit $p^z\to\infty$ in \eq{sc}, then the leading divergence in the $x\to1$ limit is
\begin{align}\label{eq:leaddiv1}
   \lim_{x\to1}\tilde q^{(1)}_{\rm cs}(x,\vec{k}_\perp,p^z\!=\!\infty) &={g^2\mu^{2\eps} C_F\over (2\pi)^{d-1}} {1\over \vec{k}_\perp^2} {2(1-x)^{2\eps}\over(1-x)}\,,
\end{align}
where the exponent $2\epsilon$ has the opposite sign to the qPDF case. After Fourier transforming to $b_\perp$ space, we obtain
\begin{align}\label{eq:leaddiv2}
   \lim_{x\to1}\tilde q^{(1)}_{\rm cs}(x,\vec{b}_\perp,p^z\!=\!\infty) &= {g^2 C_F\over 8\pi^2} \Gamma(-\eps) {2(1-x)^{2\eps}\over(1-x)}(b_\perp^2 \mu^2)^\eps\,,
\end{align}
where $b_\perp=|\vec{b}_\perp|$. Thus the leading threshold logarithm in \eq{pseudo} is reproduced after expansion in $\eps$. Since the factorization for the pPDF or spatial correlator works in the small $b_\perp$ limit, the physical scale in the threshold logarithm is proportional to $(1-x)^{-2}b_\perp^{-2}$, which approaches the UV fixed point in the $x\to1$ limit.

The different dynamical behaviors of the threshold logarithms in the qPDF and pPDF can be understood in this way: in the pPDF case, since it corresponds to the primordial TMD, the emitted gluon remains off-shell with virtual mass $-k_\perp^2$ in the limit of $x\to1$. In coordinate space, small $b_\perp$ corresponds to large $k_\perp$, so the gluon is in the UV region. Moreover, since the limit $p^z\to\infty$ has been taken first, only collinear and soft emissions with $k_\perp\to0$ are allowed, and the emission of a gluon with finite $k_\perp$ is suppressed, which explains the factor $(1-x)^{2\eps}$ in analogy to the phase-space measure.
However, in the qPDF case, since $k_\perp$ is integrated over, the limit $x\to1$ includes contributions from both hard and soft transverse momentum modes, with the latter being sensitive to IR physics. If we first take the soft transverse momentum limit of \eq{sc} at finite $p^z$ and $x$, then $k_\perp$ is bounded by both $(1-x)p^z$ and $xp^z$. Therefore, it is the integration over $k_\perp$ that changes the factor $2(1-x)^{2\eps}$ to $1+(1-x)^{-2\eps}$ in \eq{sc1}, which agrees with the fact that the probability of soft gluon emission is divergent. At the end, we note that there is a related comparison between $k_T$-resummation of the standard TMD and threshold resummation of the PDF in Refs.~\cite{Li:1998rw,Li:1998is}.

The derivations of Eqs.~(\ref{eq:sc}), (\ref{eq:sc1}) and (\ref{eq:leaddiv1}) can be found in \app{1loopsail}.\\

To make more accurate comparisons, let us calculate the full one-loop expression of $\tilde q(x,\vec{k}_\perp,p^z)$. We
first write down the spatial correlator derived in Ref.~\cite{Izubuchi:2018srq} with $b^\mu=(0,0,0,z)$, in the Lorentz covariant form,
\begin{align}\label{eq:onelooph}
	&\tilde h^{(1)}_{\gamma^t}(\lambda = -p\cdot b, -b^2\mu^2) \nn\\
	&={\alpha_sC_F\over 2\pi}\int_{0}^1 dx\ e^{-ix\lambda}\left\{ P_{qq}^{(0)}(y)\left[-{1\over\epsilon} - \ln{-b^2\mu^2e^{2\gamma_E}\over4}\right] \right. \nn\\
	&\qquad  - P_{qq}^{(0)}(y) - \left(4\ln(1-x)\over 1-x\right)_+ +2(1-x)_+ \nn\\
	&\qquad \left. + \delta(1-x)\left[ {3\over 2}\ln{-b^2\mu^2e^{2\gamma_E}\over 4} + {5\over2}\right] \right\}\,,
\end{align}
where the leading-order (LO) splitting function is given as
\begin{align}
    P_{qq}^{(0)}(x) =\left({1+x^2\over 1-x}\right)_+\,.
\end{align}
Then we use Lorentz invariance to generalize \eq{onelooph} to the case when $b^\mu=(0,\vec{b}_\perp, z)$ in Eq.\ \eqref{eq:tmd} to obtain $\tilde q(x,\vec{k}_\perp,P^z)$.

By using the formula for the Fourier transform of the logarithmic term,
\begin{align}
	&\int {d\lambda\over 2\pi} e^{ixp^z z }\ln\big(-b^2\mu^2e^{2\gamma_E}/4\big)\\
	&= \left[\ln{(\mu^2b_0^2)} + 2\Gamma(0,p^zb_\perp)\right]\delta(x)- \left({e^{-|x|p^zb_\perp} \over |x|}\right)_{\oplus}\,,\nn
\end{align}
where $b_0^2=b_\perp^2 e^{2\gamma_E}/4$, we obtain
\begin{align}\label{eq:tmd1loop}
    &\tilde q^{(1)}(x,\vec{b}_\perp,P^z)\Big/{\alpha_sC_F\over 2\pi}\\
    &= \left[P_{qq}^{(0)}(x)\left(-{1\over\epsilon}- \left[\ln{(\mu^2b_0^2)} + 2\Gamma(0,p^zb_\perp)\right] -1\right) \right.\nn\\
    &\qquad \left.- \left(4\ln(1-x)\over 1-x\right)_+  +2(1-x)_+ \right]\theta(x)\theta(1-x) \nn\\
	&+  \delta(1-x)\left({3\over2}\left[\ln{(\mu^2b_0^2)} + 2\Gamma(0,p^zb_\perp)\right] + {5\over2}\right) \nn\\
	& + \int_0^1 dy\ P_{qq}^{(0)}(y)\left({e^{-|x-y|p^zb_\perp} \over |x-y|}\right)_{\oplus} \!-\! {3\over2}\left({e^{-|1-x|p^zb_\perp} \over |1-x|}\right)_{\oplus}\,.\nn
\end{align}

As one can see, the threshold logarithms are evident at finite $b_\perp$ and $p^z$ in \eq{tmd1loop}. The leading threshold logarithms come from the overlap of collinear and soft divergences of the 3D momentum distribution as shown in \eqs{leaddiv1}{leaddiv2}. The convolution integral in the last line of \eq{tmd1loop} can affect the threshold logarithms depending on the limits in \eqs{limits1}{limits2}.

First, we have
\begin{align}
	\lim_{b_\perp\to0} \left[\ln{(\mu^2b_0^2)} + 2\Gamma(0,p^zb_\perp)\right] &= \ln{\mu^2\over 4p_z^2}\,,\\
	\lim_{p^z\to\infty} \left[\ln{(\mu^2b_0^2)} + 2\Gamma(0,p^zb_\perp)\right] &= \ln{(\mu^2b_0^2)}\,.
\end{align}

Then, by explicitly carrying out the convolution integral, we obtain
\begin{align}
	&\lim_{b_\perp\to0} \int_0^1 dy\ P_{qq}^{(0)}(y)\left({e^{-|x-y|p^zb_\perp} \over |x-y|}\right)_+ \\
	=& \left\{ \begin{array}{lc}
		\frac{2 \left(x^2+1\right) \log \left(\frac{x-1}{x}\right)+2 x+1}{2 (x-1)} & x\!>\!1\\
		\frac{-4 x^2+2 \left(x^2+5\right) \log (1-x)+2 \left(x^2+1\right) \log (x)+6 x+1}{2 (1-x)} & 0\!<\!x\!<\!1\\
		\frac{2 \left(x^2+1\right) \log \left(\frac{x-1}{x}\right)+2 x+1}{2 (1-x)} & x\!<\!0\\
	\end{array}\right.\nn
\end{align}
plus its virtual part proportional to $\delta(1-x)$. Plugging the above result into \eq{tmd1loop}, we exactly reproduce the qPDF at one-loop with its threhsold logarithms~\cite{Izubuchi:2018srq}.

On the other hand, we find that
\begin{align}
    &\lim_{p^z\to\infty} \int_0^1 dy\ P_{qq}^{(0)}(y)\left({e^{-|x-y|p^zb_\perp} \over |x-y|}\right)_+\nn\\
    &= {\cal O}\left(e^{-|1-x|p^zb_\perp}, {e^{-|1-x|p^zb_\perp}\over |1-x|p^zb_\perp}, {e^{-|x|p^zb_\perp}\over |x|p^zb_\perp}\right)\,.
\end{align}
Therefore, as long as
\begin{align}\label{eq:kin1}
    |1-x|p^zb_\perp \gg 1\,,\ \ \ \ \ |x|p^zb_\perp \gg 1\,,
\end{align}
the above convolution integral vanishes as exponentially suppressed contributions. As a result, the 3D momentum distribution is restricted to the physical region $0<x<1$, and we reproduce the threshold logarithms in the pPDF or spatial correlator in \eq{spatial}.

According to \eq{limits2}, we can identify that the limits in \eq{kin1} correspond to
\begin{align}\label{eq:kin2}
    zp^z\gg 1/(1-x)\,,\ \ \ \ \ \ zp^z\gg 1/x
\end{align}
for the spatial correlator $\tilde h_{\gamma^t}(\lambda=zp^z,-b^2=z^2)$, which implies that the threshold logarithms are sensitive to the long-range correlations.\\

Last but not least, we note that the collinear divergence in the 3D momentum distribution in \eq{tmd1loop} is the same as that in the PDF. Therefore, if $b_\perp\ll \Lambda_{\rm QCD}^{-1}$, the 3D momentum distribution should also satisfy a factorization formula that goes as
\begin{align}
    \tilde q^{(1)}(x,\vec{b}_\perp,P^z) &\!=\! \int{dy \over |y|}\ C'\Big({x\over y}, \mu^2b_0^2, yP^zb_\perp\Big) q(y,\mu)+ \ldots\,,
\end{align}
where the one-loop matching coefficient $C'$ can be obtained by subtracting the $1/\eps$ poles from \eq{tmd1loop}. 

Equivalently, the reproduction of the threshold logarithms in the appropriate limits in \eqs{limits1}{limits2} correspond to the reduction of the matching coefficient for the 3D momentum distribution to those for the qPDF and pPDF in the corresponding limits,
\begin{align}
    \lim_{b_\perp\to0} C'\Big({x\over y}, \mu^2b_0^2, p^zb_\perp\Big) &= C\Big({x\over y}, {\mu^2\over p^2_z}\Big)\,,\\
    \lim_{p^z\to\infty} C'\Big({x\over y}, \mu^2b_0^2, p^zb_\perp\Big) &= {\cal C}\Big({x\over y}, z_0^2\mu^2\Big)\,.
\end{align}

Therefore, this 3D momentum distribution can be used as an alternative lattice observable to extract the PDF. Since both $p$ and $b$ can be of any spatial direction, the data will cover a wider range of $(p\cdot b, -b^2)$ space, and we can choose $p$ to be along the diagonal directions on the lattice to reach larger boost momentum. Nevertheless, to increase the coverage of $(p\cdot b, -b^2)$ space, we also have to let $b$ be off-axis, and thus cannot construct a smooth Wilson line along the $b$ direction, which may complicate the renormalization and introduce discretization effects that break the spatial rotational symmetry. In Ref.~\cite{Musch:2010ka}, a step-like Wilson line geometry was considered, and the study there showed that the Wilson line operator is still multiplicatively renormalizable and the rotational symmetry is weakly broken with smeared gauge links.

\section{Threshold resummation at NLL accuracy}
\label{sec:nllTR}

\subsection{NLL Resummation Formula}
\label{sec:nllope}

Albeit their differences from the DIS and DY cases, the previous section demonstrated that the threshold logarithms in qPDF and pPDF indeed arise from the leading divergences of the 3D momentum distribution in the $x\to1$ limit. Now, we turn our attention to the resummation of these potentially large logarithms. The threshold resummation is most conveniently performed in the Mellin-moment space. Let us work on the OPE of the $\MS$ spatial correlator at short distance, where we write $b^2 = -z^2$ for $b^\mu = (0,0,0,z)$,
\begin{align}
    \tilde h_{\gamma^t}(\lambda, z^2\mu^2) & = \sum_{N=0}^\infty \frac{(-i \lambda )^N}{N!}C_N(\alpha_s(\mu),z^2_0\mu^2) a_N(\mu) \!+ \ldots \,,\label{eq:ope}
\end{align}
where $a_N(\mu)$ is the $(N+1)$-th moment of the PDF,
\begin{align}
    a_N(\mu) &=\int_{-1}^1 dy~y^{N}q(y,\mu) \,.\label{eq:moment}
\end{align}
At NLO, the Wilson coefficient $C_N$ is given by
\begin{align}
C_N^{\rm NLO} &= \int_0^1 dw\ w^N \mathcal{C}^{\rm NLO}(w,z^2\mu^2)\\
&= \frac{\alpha_s(\mu) C_F}{2\pi} \biggl[ \Bigl(
\frac{3\!+\!2N}{2\!+\!3N\!+\!N^2}
\!+\! 2H_N\Bigr) \ln(z_0^2\mu^2)
\nn\\
&\quad\quad
+   \frac{5\!+\!2N}{2\!+\!3N\!+\!N^2}+2(1-H_N)H_N - 2 H_N^{(2)}\biggr]\,,\nn
\end{align}
where Harmonic numbers are defined as $H_N=\sum_{i=1}^N 1/i$ and $H_N^{(2)} =\sum_{i=1}^N 1/i^2$ and $\mathcal{C}^{\rm NLO}$ is the matching coefficient in Eq.\ \eqref{eq:io-Q-fact} at NLO.

The threshold limit corresponds to $N\to\infty$, given as
\begin{align} \label{eq:cn1}
    \lim_{N\to\infty}C_N^{\rm NLO} &= \frac{\alpha_s(\mu) C_F}{2\pi} \biggl[2\ln N' \ln(z_0^2\mu^2) \nn\\
    &\qquad - 2\ln^2 N' + 2 \ln N' - {\pi^2\over3}\biggr]\,,
\end{align}
where $N'=N e^{\gamma_E}$. As one can see, the leading and sub-leading threshold logarithms are, respectively,
\begin{align}
    \alpha_s \ln^2 N'\,,\ \ \ \ \alpha_s \ln N'\,.
\end{align}
We propose to use the standard technique of threshold resummation to resum these large logarithms~\cite{Catani:1996yz}.
Usually, it requires resummation at NLL accuracy to guarantee the convergence when one performs the inverse Mellin transform to obtain the resummed matching coefficient $C$ or $\cal C$~\cite{Catani:1996yz}.

From now on let us denote $a_s= \alpha_s/(2\pi)$. The all-order resummed form of the Wilson coefficient $C_N$ inspired by the standard threshold resummation technique is
\begin{align}\label{eq:nll}
\ln C_N^{\rm {NLL}} & = \int dx {x^{N-1}-1\over 1-x} \biggl[\int_{\mu^2}^{(1-x)^{a}\over z_0^2}{dk^2\over k^2 }A(\alpha_s(k^2))\nn\\
&\qquad \qquad  + B(\alpha_s((1-x)^a/z_0^2))\biggr]\,,
\end{align}
where $a=-2$ as it corresponds to the logarithm $\ln ((1-w)^2z^2_0\mu^2)$ in \eq{pseudo}. As a consequence of the negative $a$, $k^2$ is evolved to the UV fixed point in the limit $x\to1$. In comparison, $a=2$ and $1$ correspond to threshold resummation for DIS and DY cases, respectively~\cite{Moch:2005ba}, which reaches the Landau pole. Additionally,
\begin{align}
	A(\alpha_s) &= A^{(0)} a_s + A^{(1)} a_s^2 + \ldots\,,\\
	B(\alpha_s) &= B^{(0)} a_s + B^{(1)} a_s^2 + \ldots\,.
\end{align}
Since the leading threshold logarithm arises from the overlap of collinear and soft divergences in the 3D momentum distribution, which is the common origin of Sudakov double logarithms~\cite{Sudakov:1954sw, Contopanagos:1996nh}, the coefficient $A$ should be the universal cusp anomalous dimension~\cite{Korchemsky:1987wg,Becher:2009cu}. Based on the one-loop Wilson coefficient in \eq{cn1},
\begin{align}
A^{(0)} &= -B^{(0)} = 2C_F\,.
\end{align}
For NLL resummation which neglects ${\cal O}(\alpha_s^2\ln N')$ terms, we only need $A^{(1)}$, the two-loop cusp anomalous dimension~\cite{Korchemsky:1987wg},
\begin{align}\label{eq:a1}
	A^{(1)}=2C_F\left[C_A\left({67\over18}-{\pi^2\over6}\right) - {10\over9}n_fT_F\right]\,,
\end{align}
where $C_A=3$, $T_F=1/2$, and $n_f$ is the number of active quark flavors.

According to the NNLO matching coefficient ${\cal C}$ calculated in Ref.~\cite{Li:2020xml},
\begin{align}
&\lim_{N\to\infty}C_N^{\rm NNLO}(\alpha_s(z_0^{-1}),1) \\
&= a_s^2 \left[C_F^2 \left(2\ln^4N' - 4 \ln^3N' + \big(2+{2\pi^2\over 3}\big)\ln^2N'\right)\right.\nn\\
& + C_FC_A \left({22\over9}\ln^3N' - 2\left({67\over18} - {\pi^2\over 6} + {11\over 6}\right)\ln^2N' \right)\nn\\
&\left. + C_Fn_fT_F \left( - {8\over 9}\ln^3N' + {32\over 9} \ln^2N' \right)\right] + {\cal O}(a_s^2\ln N')\,,\nn
\end{align}
which exactly verifies \eq{a1} and the expansion of \eq{nll} to NNLO, especially that $a=-2$.

Plugging the above results into \eq{nll}, we obtain the resummed Wilson coefficient,
\begin{align}\label{eq:NLLkernel}
\ln C_N^{\rm NLL} (\alpha_s(\mu),z^2_0\mu^2) & = -{\pi^2\over 3} a_sC_F + \ln N' g_1(\tau,L ) \nn\\
&\qquad\qquad + g_2(\tau,L),
\end{align}
where $\tau = \beta_0 a_s \ln N'$, $L=\ln(z_0^2\mu^2)$. The first term is of ${\cal O}(a_s)$, so it can be either exponentiated or kept in the NLO matching coefficient for NLL resummation. Here, we exponentiate it in accordance to the treatment in the ASV analysis~\cite{Aicher:2010cb}. The functions $g_1$ and $g_2$ are
\begin{align}
&g_1(\tau,L)\label{eq:g1} \\
=& -{A^{(0)}\over 2\beta_0\tau} \left[-2 \tau -\left(1-\frac{\tau  L}{\ln N' }\right)\ln \left(1-\frac{\tau  L}{\ln N' }\right)\right.\nn\\
&\left.+\left(1+\tau  \left(2-\frac{L}{\ln N' }\right)\right)\ln \left(1+\tau  \left(2-\frac{L}{\ln N' }\right)\right) \right]\,,\nn\\
&g_2(\tau,L)\label{eq:g2} \\
=& -{A^{(0)}\beta_1\over 4\beta_0^3}\left[\ln^2\left(1+\tau  \left(2\!-\!\frac{L}{\ln N' }\right)\right)-\ln^2\left(1\!-\!\frac{\tau  L}{\ln N' }\right)\right]\nn\\
&-\left({A^{(0)}\over \beta_0} \!-\! {A^{(1)}\over \beta_1}\right){\beta_1\over 2\beta_0^2} \biggl[-2 \tau +\ln \left(1+\tau  \left(2\!-\!\frac{L}{\ln N' }\right)\right)\nn\\
&- \ln \left(1-\frac{\tau  L}{\ln N' }\right)\biggr]\nn\\
&- {B^{(0)}\over 2\beta_0} \biggl[\ln \left(1+\tau  \left(2-\frac{L}{\ln N' }\right)\right)- \ln \left(1-\frac{\lambda  L}{\ln N' }\right)\biggr] \,,\nn
\end{align}
where the $\beta_0=(33-2n_f)/6$ and $\beta_1=(153-19n_f)/6$ are the lowest coefficients in the QCD $\beta$-function
\begin{align}
    \beta(a_s) &= {da_s(\mu)\over d\ln \mu^2} = - (\beta_0 a_s^2 + \beta_1 a_s^3 + \ldots)\,.
\end{align}

Since $\tau\sim 1$ at NLL accuracy and $L$ is not resummed at the fixed order, \eqs{g1}{g2} correspond to the hierarchy $L\ll \ln N'$. For practical applications to lattice QCD calculations, the range of available $z$ is not necessarily small, so one might also need to include the DGLAP evolution~\cite{Altarelli:1977zs,Dokshitzer:1977sg,Gribov:1972ri} of the PDF or Mellin moments to resum $L$. To accomplish this, we start from the evolution equation for the Wilson coefficient $C_N$,
\begin{align}\label{eq:rge}
    \Big[{\partial \over \partial \ln \mu^2} + \beta(a_s(\mu)){\partial \over \partial a_s} - \gamma_N\Big] C_N=0\,,
\end{align}
where $\gamma_N$ is the anomalous dimension which expands in $a_s$,
\begin{align}
    \gamma_N &= \gamma^{(0)}_N a_s + \gamma^{(1)}_N a_s^2 + \ldots\,.
\end{align}
At one-loop~\cite{Izubuchi:2018srq},
\begin{align}
    \gamma_N^{(0)} &= \left[\frac{3\!+\!2N}{2\!+\!3N\!+\!N^2} \!+\! 2H_N\right]C_F\,.
\end{align}
Therefore, the solution to \eq{rge} at leading logarithmic (LL) accuracy is
\begin{align}
\label{eq:CNevo}
C_N^{\rm evo}(\alpha_s(\mu),z_0^2\mu^2) &\!=\! C_N \big(\alpha_s(z_0^{-1}),1\big)\!\left(\!\frac{\alpha_s(z_0^{-1})}{\alpha_s(\mu)}\!\right)^{\!\frac{\gamma_N^{(0)}}{\beta_0}}\,,
\end{align}
where the superscript ``evo'' indicates the DGLAP-evolved Wilson coefficient.

Then, we can perform threshold resummation of the Wilson coefficient at $\mu = z_0^{-1}$, $C_N \big(\alpha_s(z_0^{-1}),1\big)$, which is related to \eq{NLLkernel} by setting $L=0$,
\begin{align}\label{eq:NLLkernel2}
\ln C_N^{\rm NLL}\big(\alpha_s(z_0^{-1}),1\big) & = -{\pi^2\over 3} a_sC_F + \ln N' g_1(\tau ,0) \nn\\
&\qquad\qquad + g_2(\tau,0)\,,
\end{align}
where
\begin{align}
g_1(\tau ,0) &=  -{A^{(0)}\over 2\beta_0\tau} \left[-2\tau + (1+2\tau)\ln(1+2\tau) \right]\,,\label{eq:g1b} \\
g_2(\tau ,0) &= -{A^{(0)}\over \beta_0} {\beta_1\over 4\beta_0^2}\ln^2(1+2\tau)  - \left({A^{(0)}\over \beta_0}- {A^{(1)}\over \beta_1}\right){\beta_1\over 2\beta_0^2}\nn\\
&\quad \times \big[\!-\!2\tau \!+\! \ln(1+2\tau)\big]- {B^{(0)}\over 2\beta_0} \ln(1+2\tau) \,.\label{eq:g2b}
\end{align}

Using the inverse Mellin transform, we can eventually obtain the resummed matching coefficient with LL DGLAP evolution as
\begin{align}
\label{eq:final}
&{\cal C}^{\rm NLL+evo}(w,z^2\mu^2)=e^{-{\pi^2\over 3}a_sC_F}  {1\over 2\pi i} \int_{C-i\infty}^{C+i\infty}dN\ w^{-N}\nn\\
&\qquad\times \exp\Big[ \ln N' g_1(\tau,0) + g_2(\tau,0) \Big]\left(\!\frac{\alpha_s(z_0^{-1})}{\alpha_s(\mu)}\!\right)^{\!\frac{\gamma_N^{(0)}}{\beta_0}} \,,
\end{align}
which can be carried out numerically. The resummed matching coefficient $C^{\rm {NLL+evo}}\left(\xi, \mu/p^z\right)$ for the qPDF is related to ${\cal C}^{\rm NLL+evo}(w,z^2\mu^2)$ through a double Fourier transform~\cite{Izubuchi:2018srq}, which becomes more involved and is beyond the scope of this work. Notably, both $g_1(\tau ,0)$ and $g_2(\tau ,0)$ are regular for all $N$. Therefore, unlike DIS and DY, there is no Landau pole along the path of integration in \eq{final}, which is due to the UV fixed point that we mentioned above.\\

In the previous work by some of the authors~\cite{Gao:2020ito}, the lattice data has been analyzed with the NLO Wilson coefficients $C^{\rm NLO}_N(\alpha_s(\mu),z^2_0\mu^2)$ in Eq.\ \eqref{eq:ope}. Now we can include the threshold resummation at NLL and DGLAP evolution by replacing $C_N$ in \eq{ope} with $C_N^{\rm evo}$ from \eq{CNevo} then further replacing $C_N$ at $\mu = 1/z_0$ in \eq{CNevo} with $C_N^{\rm NLL}$ in \eq{NLLkernel2}. 
We can also combine the fixed-order and threshold resummation for all $N$ by replacing $C_N$ with
\begin{align}
    C_N^{\rm NLO+NLL} &= C_N^{\rm NLO} + C_N^{\rm NLL}  - C_N^{\rm NLOexp}\,,
\end{align}
where $C_N^{\rm NLOexp}$ is $C_N^{\rm NLO}$ expanded to ${\cal O}(a_s)$, which is the same as the right-hand side of \eq{cn1},
so that at small $N$ the latter reduces to $C_N^{\rm NLO}$ up to ${\cal O}(a_s^2\ln^2 N',a_s^2\ln N',a_s^2)$ corrections. Note NLL threshold resummation is accompanied by the NLO Wilson coefficient.

\subsection{Estimating impacts of NLL resummation on the large-$x$ exponent}
\label{sec:largex}

Since the resummed Wilson coefficient is generally greater than the fixed-order one at large $N$, the fitted higher moments with threshold resummation should be smaller, which will suppress the PDF at large $x$ or lead to a larger value of $\beta$.

For a more quantitative estimate of the effect of threshold resummation on $\beta$, we choose a PDF parametrized as the form
\begin{align}
\label{eq:beta}
    f(x)&= {\cal N}x^\alpha (1-x)^\beta {\cal G}(x)\,,
\end{align}
where ${\cal N}$ is a normalization constant and ${\cal G}(x)$ is a smooth and well-behaved function within $0<x<1$. It was shown~\cite{Gao:2020ito} that at large $N$
\begin{align}
    \langle x^N \rangle \propto N^{-\beta -1} \left[1 + {\cal O}(1/N)\right] \,,
\end{align}
regardless of the function ${\cal G}(x)$. If $\langle x^N \rangle$ is known, then $\beta$ can be approximated as
\begin{align}
    \beta &= - {d\ln \langle x^N \rangle \over \ln N} -1 + {\cal O}(1/N)\,.
\end{align}
For given lattice matrix elements, if one solely changes the matching coefficient from $C_N^{\rm NLO}$ to $C_N^{\rm NLO+NLL}$, then $\beta$ will be changed by
\begin{align}
\label{eq:dbeta}
    \delta \beta(N) &= \beta^{\rm NLO+NLL} - \beta^{\rm NLO}\\
    &= {d\over d\ln N}\left(\ln C_N^{\rm NLO+NLL}-\ln C_N^{\rm NLO}\right) + {\cal O}(1/N^2)\,.\nn
\end{align}

\begin{figure}
\centering
	\includegraphics[width=0.9\linewidth]{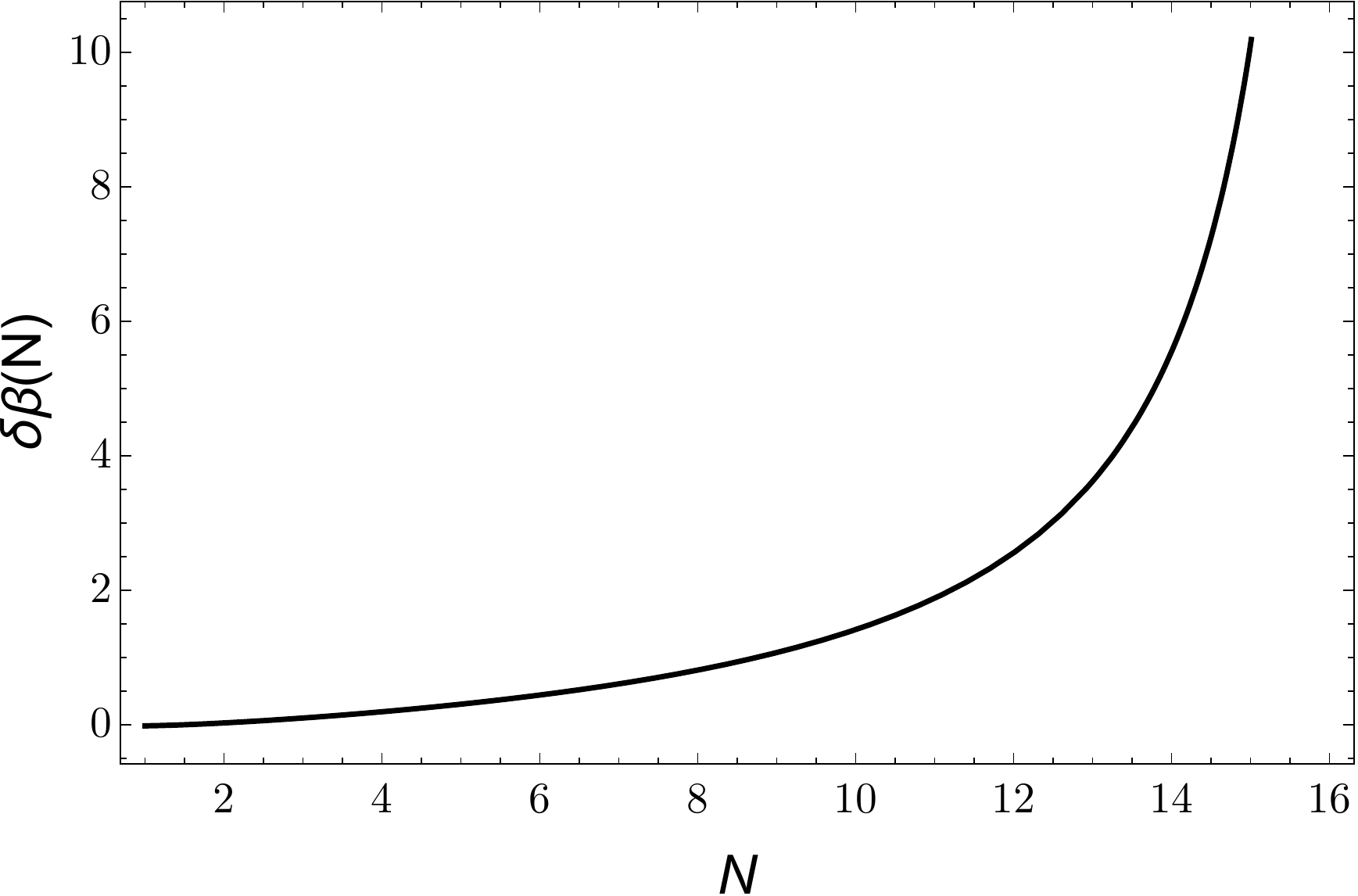}
	\caption{Estimated change $\delta\beta$ of the large-$x$ exponent of PDF (c.f. Eq.\ \eqref{eq:beta} and Eq.\ \eqref{eq:dbeta}) as a function of the order of the Mellin moment when the matching coefficient is changed from $C_N^{\rm NLO}$ to $C_N^{\rm NLO+NLL}$ with $\mu=1/z_0=3.2$ GeV.}
	\label{fig:beta}
\end{figure}

For the purpose of this discussion we do not include DGLAP evolution, that is, we set $\mu=1/z_0$ so that $\ln (z_0^2\mu^2)=0$ in $C_N^{\rm NLO+NLL}$ and $C_N^{\rm NLO}$. In \fig{beta}, we plot $\delta \beta(N)$ as a function of $N$ for $\mu=3.2$~GeV and $\alpha_s(\mu)=0.242$, within the range of $N \in [1, 16]$. We have to truncate at $N=16$ because $C^{\rm NLO}_N$ would become negative at $N>16$. At $N=4$, threshold resummation increases the NLO fit of $\beta$ by about $0.2$, which is a weak effect. As $N$ increases the impact of threshold resummation becomes stronger. The increment of $\beta$ by threshold resummation was also observed in the ASV fit~\cite{Aicher:2010cb} of the DY data. Since $C_N^{\rm NLO}$ falls close to zero as $N$ increases, it will significantly enhance the fitted higher moments and lead to smaller and even negative $\beta$, which could be compensated by the rapid growth of $\delta \beta$ shown in \fig{beta}. This indicates that threshold resummation is necessary and may help stabilize the fitting of $\beta$.

\section{Application to lattice QCD calculations}
\label{sec:num}

\begin{table}
\centering
\begin{tabular}{ c | c | c  c }
\hline \hline
    \multirow{2}{*}{$n_z$} &  \multicolumn{2}{c}{$P^z$ (GeV)} \\
    \cline{2-3}
    & $a=0.06$ fm & $a=0.04$ fm \cr
\hline
    0  &   0  & 0 \\
    1  &  0.43 & 0.48 \\
    2  &  0.86 & 0.97  \\
    3  &  1.29 & 1.45  \\
    4  &  1.72 & 1.93  \\
    5  &  2.15 & 2.42 \\
\hline \hline
\end{tabular}
\caption{Boost momenta, $P^z$(GeV), of pion used in this work for the two different lattice spacings~\cite{Gao:2020ito}.}
\label{tbl:pzlist}
\end{table}

In this section, we study the impact of threshold resummation on the valence PDF of pion based on the recent lattice QCD results of Ref.~\cite{Gao:2020ito}. Since our resummation has been derived in the Mellin-moment space, it is naturally implemented on the coordinate-space lattice data. Using the OPE formula in \eq{ope} with $C_N^{\rm NLO}$, $C_N^{\rm NLO+NLL}$ with DGLAP evolution, and $C_N^{\rm NLOevo}$, the coefficient with only DGLAP evolution included, we study their effects by fitting the data with a finite number of moments and also a parameterization model for the pion valence PDF.

The lattice results we use are from a previous publication by some of the authors~\cite{Gao:2020ito}. These results are for a pion wtih $300$~MeV mass, including two different lattices with spacing $a = 0.04$~fm and $0.06$~fm, pion momenta $P^z=(2\pi n_z)/L_s$ with $n_z$ from 0 to 5 listed in \tbl{pzlist}, and volumes $(2.56)^3$~fm$^3$ and $(2.88)^3$~fm$^3$ for each $a$ respectively. Therefore, the largest boost momentum on the finer lattice is $P^z=2.42$~GeV. 

\subsection{Model-independent moments}
\label{sec:moments}

In the OPE of the spatial correlator, the higher-twist contributions and higher moments contributions are suppressed by ${\cal O}(z^2\Lambda_{\rm QCD}^2)$ and ${\cal O}(\lambda^N/N!)$, respectively. The former restricts $z$ to be short, which in turn limits the range of $\lambda=zP^z$ with finite $P^z$. While the higher moments $a_N$ are further suppressed by the asymptotic behavior of PDF at large $x$, the currently available lattice results are only sensitive to the lowest ones. In Ref.~\cite{Gao:2020ito} it was found that these lattice results are not sensitive to moments beyond $N=6$. Furthermore, since the pion matrix elements for the valence PDF are real and symmetric in $z$, only even moments contribute. Thus, for this OPE-type analysis, reasonable signals are expected for only $\langle x^2\rangle$ and $\langle x^4 \rangle$.

The lattice data we fit to are renormalization-group-invariant ratios constructed from bare matrix elements $\tilde h_{\gamma^t}(\lambda = zP^z,z^2)$~\cite{Gao:2020ito} with different pion momenta,
\begin{align}
    {\cal M}(z,P^z,P^z_0)&\equiv\frac{\tilde h_{\gamma^t}(\lambda = zP^z,z^2)}{\tilde h_{\gamma^t}(\lambda = zP^z_0,z^2)}\,,
\end{align}
which was first proposed in Ref.~\cite{Orginos:2017kos} with $P^z_0=0$ and then adapted in Ref.~\cite{Fan:2020nzz} to use $P^z_0\gg \Lambda_{\rm QCD}$ to suppress higher-twist effects. For this analysis, we choose $aP^z_0=2\pi n_z^0/L_s$ with $n_z^0=1,2$ for both the lattice spacings. The leading-twist approximation to the above ratio is
\begin{align}\label{eq:ratiotmc}
{\cal M}(z,P^z,P^z_0)&=\frac{\sum_N c_N(z^2\mu^2)\frac{(-izP^z)^N}{N!} \langle x^N \rangle_{P^z}}{\sum_N c_N(z^2\mu^2)\frac{(-izP^z_0)^N}{N!} \langle x^N \rangle_{P^z_0}}\,,
\end{align}
where $c_N(z^2\mu^2)$ is the $\overline{\textup{MS}}$ Wilson coefficient $C_{N}$ normalized by $C_0$ without changing the outcome.
We consider $c_N$ for NLO without any DGLAP evolution (denoted as NLO), NLO with LL DGLAP evolution (denoted as NLOevo), and NLO+NLL with LL DGLAP evolution (denoted as NLOevo+NLL), respectively. We use one-loop QCD $\beta$ function for the DGLAP evolution with $n_f=3$ to be consistent with the 2+1 flavor lattice ensemble. Here, $\langle x^N \rangle_{P^z}$ is the Mellin moment with target mass correction included~\cite{Chen:2016utp},
\begin{align}
\langle x^N \rangle_{P^z} &=  \langle x^N \rangle \sum_{k=0}^{N/2}\frac{(N-k)!}{k!(N-2k)!}\Big(\frac{m^2_\pi}{4P_z^2}\Big)^k\,.
\end{align}
The target-mass corrections have little effect for the pion because of the smallness of its mass $m_\pi= 300$~MeV in our case. In the previous work~\cite{Gao:2020ito}, the data has been analyzed with only the NLO Wilson coefficients. 

\begin{figure}
\centering
    \includegraphics[width=0.95\linewidth]{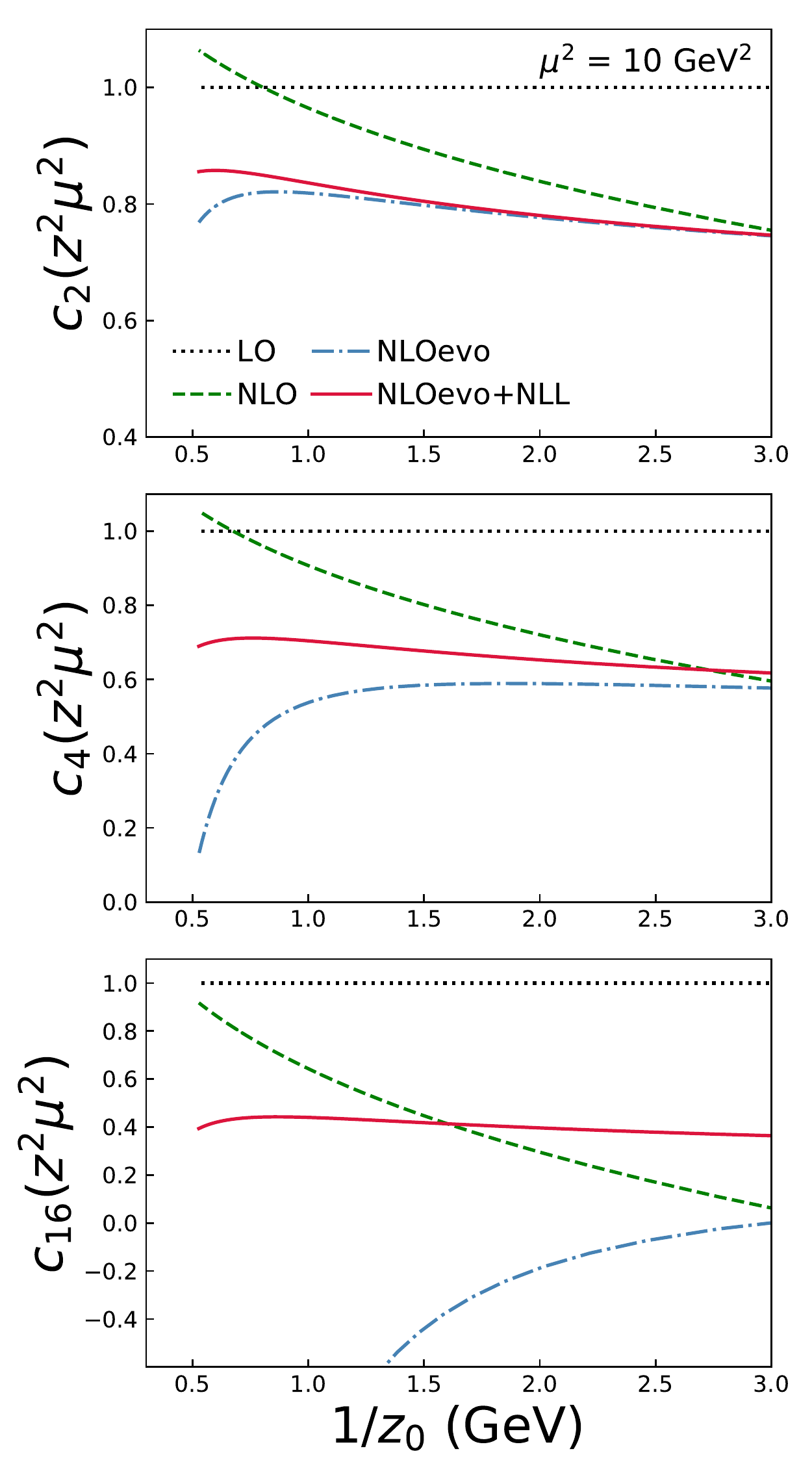}
	\caption{The Wilson coefficient $c_N(z^2\mu^2)$ with $N=2,4,16$ at LO, NLO, NLOevo, and NLOevo+NLL accuracy are shown. \label{fig:kernels}}
\end{figure}

\begin{figure}
\centering
    \includegraphics[width=0.95\linewidth]{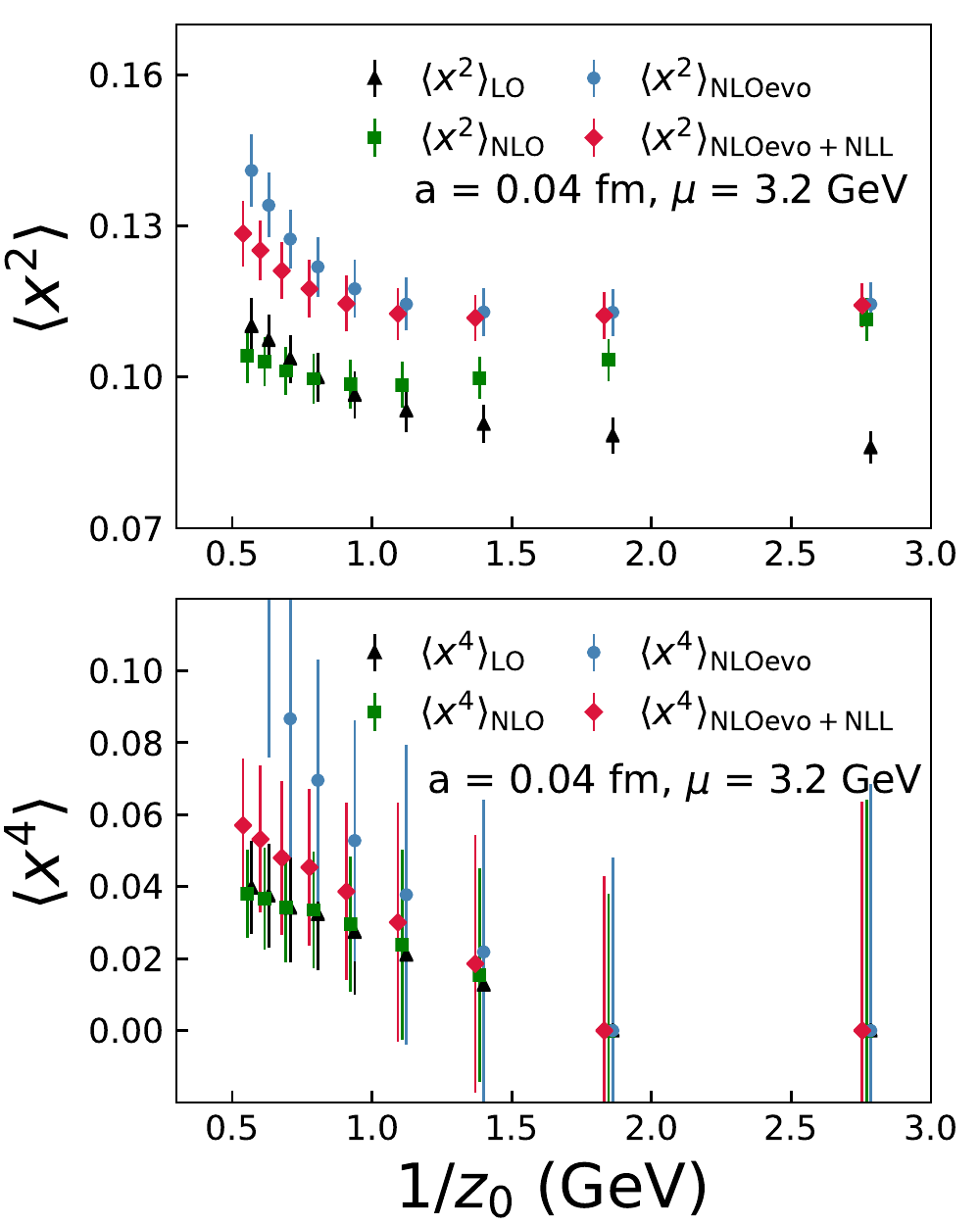}
	\caption{Extracted values of $\langle x^2\rangle$ (upper panel) and $\langle x^4\rangle$ (lower panel) by fitting $a=0.04$~fm lattice results for $n_z>n_z^0=1$ to Eq.~(\ref{eq:ratiotmc}) at each value $z$, using Wilson coefficients of different accuracy. \label{fig:mom2evo}}
\end{figure}

\begin{figure}
\centering
	\includegraphics[width=0.95\linewidth]{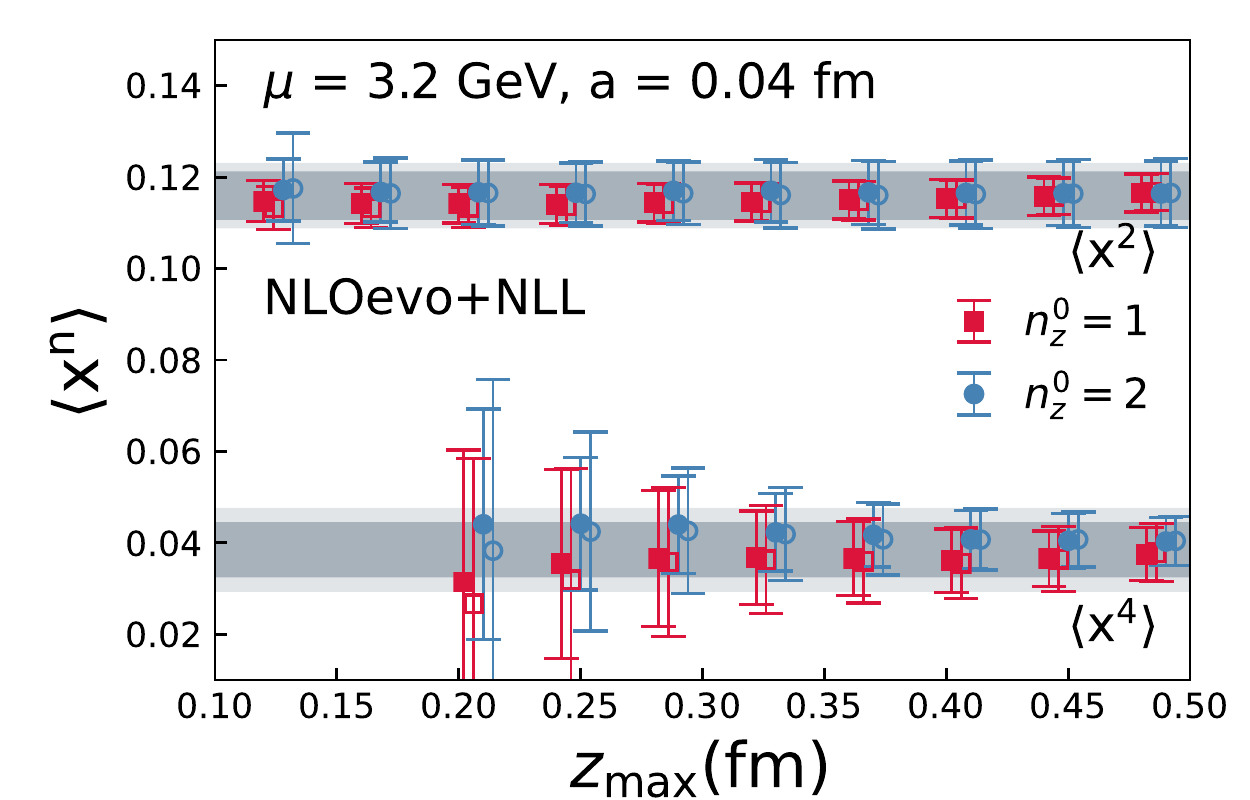}
	\includegraphics[width=0.95\linewidth]{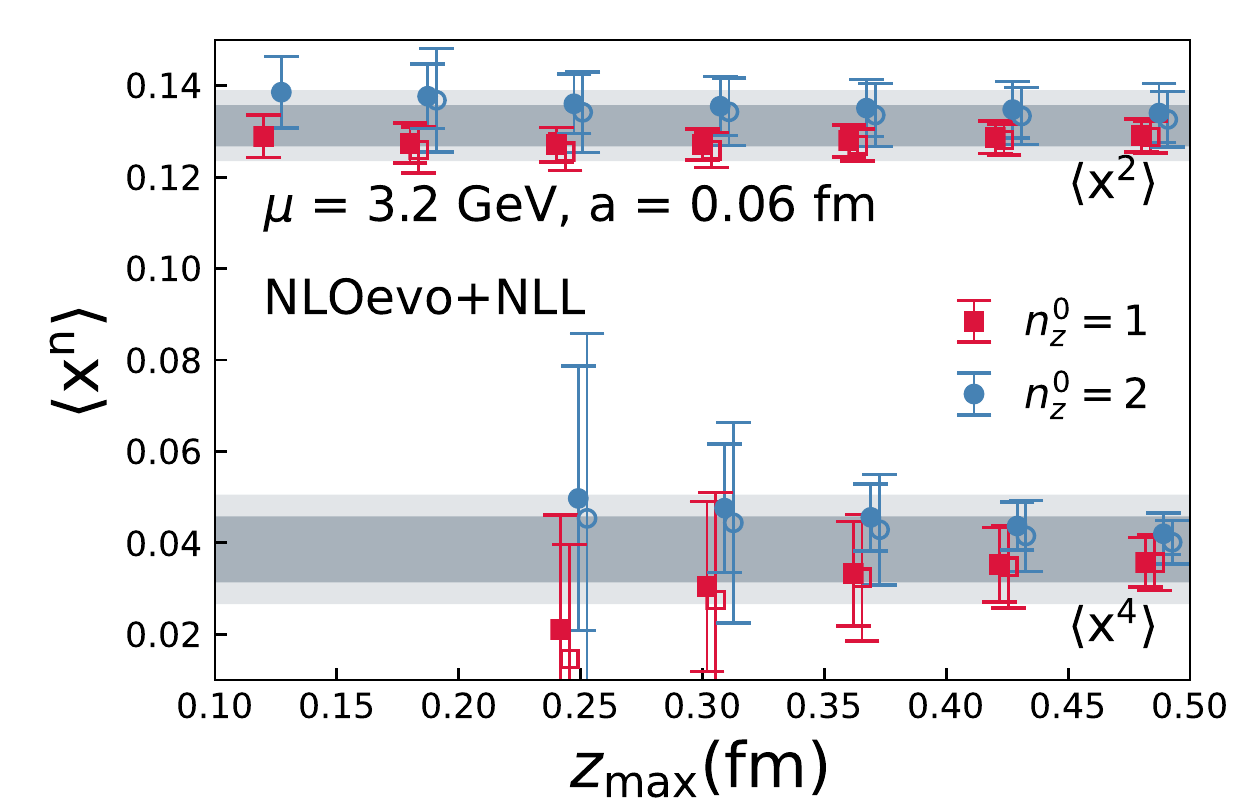}
	\caption{The combined fit results of $\langle x^2 \rangle$ and $\langle x^4 \rangle$ from the region $[z_{\rm min}, z_{\rm max}]$, plotted as a function of $z_{\rm max}$ with $z_{\rm min} = 2a$ (filled symbols) and 3a (open symbols) for a = 0.04 fm (up) and a = 0.06 fm (down).\label{fig:momfit}}
\end{figure}

In \fig{kernels}, we show $c_N(z^2\mu^2)$, with $N=2,4,16$ and at a scale $\mu =3.2$~GeV, as a function of $1/z_0$ for the leading order (LO) in $\alpha_s$, NLO, NLOevo and NLOevo+NLL. The DGLAP evolution is included here because for a large range of $z$, the logarithm $L=\ln(z_0^2\mu^2)$ can become large and reduce the predictive power of OPE with fixed-order Wilson coefficients. As one can see, the DGLAP evolution is indeed an important effect within the range of $z$ we considered, and the LL resummation of $L$ makes the Wilson coefficients change slower than the NLO ones as $1/z_0$ is varied. However, one can also notice that as $N$ increases the size of one-loop correction grows in the NLOevo Wilson coefficient, which even makes it turn negative at small scales of $1/z_0$. This is exactly caused by the large logarithmic term $-\alpha_s\ln^2 N'$ in~\eq{cn1}, especially with large $\alpha_s(z_0^{-1})$, and it implies the necessity of threshold resummation for either large $\alpha_s$ or large $N$ appearing in the Wilson coefficients. As shown in \fig{kernels}, significant impact can be observed for large $\alpha_s$ (small $1/z_0$) or $N$ when we compare the NLOevo+NLL coefficients to the NLOevo ones, though there is only mild difference for $N=2,4$ at $1/z_0 > 2$ GeV.

As mentioned above, current lattice data is only sensitive to the first few lowest moments. In \fig{mom2evo}, we take the $a = 0.04$ fm lattice data as an example, and plot the fitted second moment $\langle x^2\rangle$ and fourth moment $\langle x^4\rangle$ at each single $1/z_0$. In the plots, we skip the data point at $z=a$ (or $1/z_0=$5.5 GeV) to avoid the possible lattice discretization effects. In principle, the result $\langle x^n\rangle (\mu)$ should only depend on the $\MS$ scale $\mu$, as the $\ln(z^2)$-dependence of lattice data should be canceled by that of the Wilson coefficient. The cancellation is expected to work better at smaller $z$ or at larger $1/z_0$ since $\alpha_s(1/z_0)$ is smaller. Focusing on the second moment $\langle x^2 \rangle$, at LO one can clearly observe the $z$-dependence of $\langle x^2 \rangle_{\textup{LO}}(z)$, which grows faster at smaller scales in $1/z_0$. Beyond LO, one can find a plateau for $\langle x^2 \rangle$ which indicates that the coefficients $c_2(z^2\mu^2)$ can explain the evolution of the moments well in the plateau region. The fixed-order analysis can be found in Ref.~\cite{Gao:2020ito}. After introducing the DGLAP evolution, one can observe the plateau for $\langle x^2 \rangle_{\textup{NLOevo}}$ when $1/z_0~\gtrsim 0.8$ GeV. Beyond $1/z_0 \sim 0.8$ GeV, the plateau remains stable up to around $1/z_0\sim 3$ GeV.

For $\langle x^2 \rangle_{\textup{NLOevo+NLL}}$ one can see that threshold resummation slightly improves predictive power, as the plateau region is extended to $1/z_0~\gtrsim$ 0.6 GeV within the errors. Nevertheless, the plateau value remains almost unchanged compared to $\langle x^2 \rangle_{\textup{NLOevo}}$ as we expect. The lower scale region is not showing a good plateau but is not of concern because $\alpha_s(1/z_0)$ becomes large and is accompanied by large theoretical uncertainty. As for the fourth moment $\langle x^4 \rangle(z)$ extracted from each $z$, though noisy, one can observe the problematic behaivor of the $\langle x^4 \rangle_{\textup{NLOevo}}$ at small scale is caused by large $-\alpha_s\ln^2 N'$ in the Wilson coefficients as discussed above, and the situation is improved for $\langle x^4\rangle_{\textup{NLOevo+NLL}}$ with NLL resummation within the errors.

To stabilize the fit and extract higher moments, we then use NLOevo+NLL Wilson coefficients to perform a combined fit to all $z$ within the range [$z_{\rm min}$, $z_{\rm max}$]. We include the theoretical uncertainty in the fit by varying the factorization scale $\mu$ up and down by a factor of two as in Ref.~\cite{Gao:2020ito}. The fitted second and fourth moments are shown in \fig{momfit} as a function of $z_{\rm max}$ with $n_z^0=1,2$ and $z_{\rm min}=2a,~3a$ for $a=0.04$~fm (up) and 0.06 fm (down). We skipped $z_{\rm min}=a$ to avoid possible discretization effects. We  observed that the $z_{\rm min}$ as well as the $n_z^0$ dependence are mild within the errors. Therefore, we estimate the statistic and systematic errors as the darker and lighter bands  using the same strategy as in Ref.~\cite{Gao:2020ito}: for each bootstrap sample, we evaluate the average and standard deviation of observable $A$ from a range of $n_z^0$, $z_{\textup{min}}$ and $z_{\textup{max}}$ as ${\textup{Mean}(A)}$ and $\textup{SD}(A)$. Then we can get the statistic errors of $\overline{\textup{Mean}(A)}$, and take $\overline{\textup{SD}(A)}$ as the systematic error. In this analysis, we use $n_z^0$ = 1,2, $z_{\rm min}=2a,~3a$ and $z_{\rm max}\in[0.36, 0.48]$ fm. One can observe a plateau for both $\langle x^2 \rangle$ and $\langle x^4 \rangle$ within the errors.

\begin{figure}
\centering
	\includegraphics[width=0.95\linewidth]{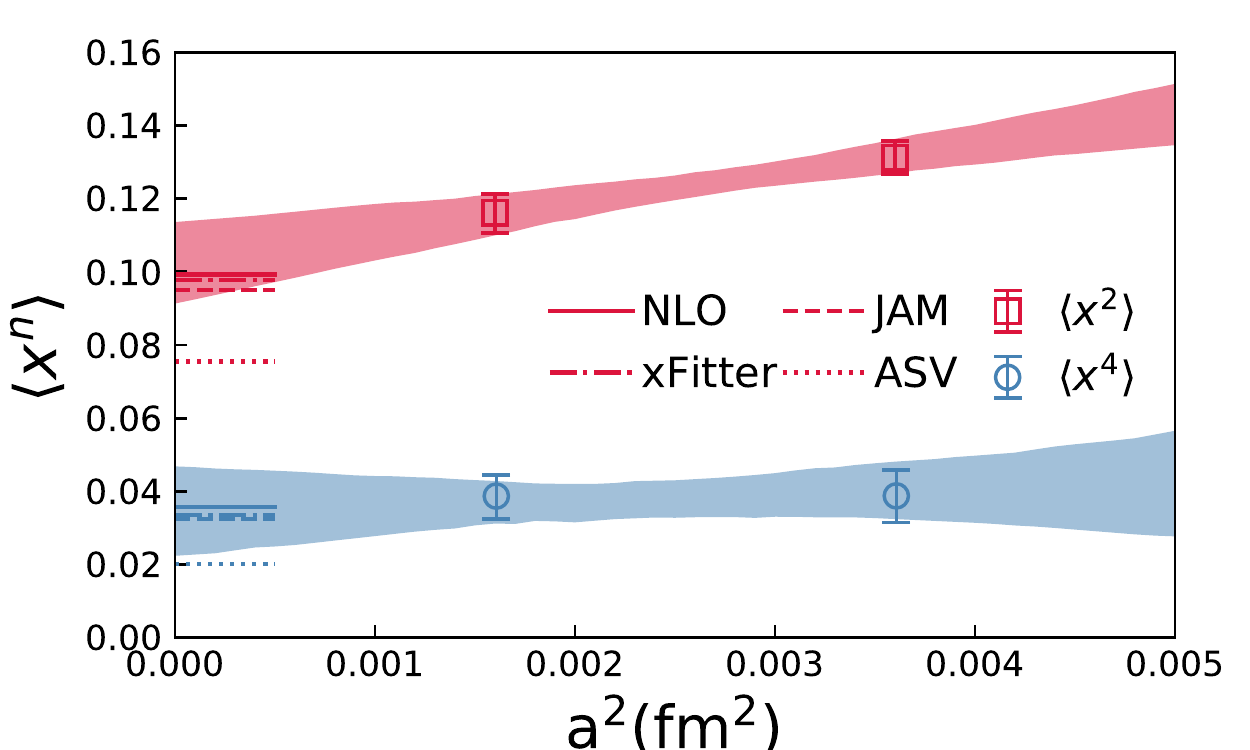}
	\caption{Continuum extrapolation of $\langle x^2 \rangle$ and $\langle x^4 \rangle$ from NLOevo+NLL fit at $\mu$ = 3.2 GeV are shown as the bands. Instead of directly extrapolating the moments, the continuum extrapolation is  performed by insert \eq{momsconti} into expression \eq{ratiotmc}. For comparisons, the moments from NLO analysis~\cite{Gao:2020ito}, {\UrlFont xFitter}~\cite{Novikov:2020snp}, JAM~\cite{Barry:2018ort} and ASV~\cite{Aicher:2010cb} valence PDF of pion are also shown as the lines. \label{fig:momsconti}}
\end{figure}

Since mild lattice spacings dependence of the moments can be observed in \fig{momfit}, a joint fit is performed for a continuum extrapolation by modifying the moments inserted in \eq{ratiotmc} as
\begin{align}\label{eq:momsconti}
&\langle x^N \rangle_{P^z} \!\to\! \big(\langle x^N \rangle \!+\! d_N a^2\big) \sum_{k=0}^{N/2}\frac{(N-k)!}{k!(N-2k)!}\Big(\frac{m^2}{4P_z^2}\Big)^k\,.
\end{align}
The continuum extrapolations are shown in \fig{momsconti}. One observes that the second moment indeed shows some lattice spacing dependence, while the fourth moment shows little dependence within errors. For comparison, we also show the moments of our previous fixed-order estimate, and the {\UrlFont xFitter}~\cite{Novikov:2020snp}, JAM~\cite{Barry:2018ort} ASV~\cite{Aicher:2010cb} pion valence PDF, which are extracted from fits with data from experiments. Though the mean value of the second moment $\langle x^2 \rangle$ = 0.103(11)(3) is slightly higher than the NLO fit 0.099(7)(5) in our previous work~\cite{Gao:2020ito}, our result still shows better agreement with {\UrlFont xFitter} and JAM. The fourth moment $\langle x^4 \rangle$ shows overall consistency with the NLO fit~\cite{Gao:2020ito} and the results extracted from experimental measurements
%experimental results 
because of the large errors. This mild change of the first few moments compared to the NLO analysis is mainly driven by the DGLAP evolution, while the threshold resummation has little impact but helps to stabilize the fit in the low scale region of $1/z_0$. To access higher moments, where the threshold resummation generally has a considerable impact, one has to increase the pion momentum while working at a small $z$. The fit results of the moments are summarized in \tbl{conti}.

\subsection{Model fit}
\label{sec:model}

Compared to the DIS and DY cross sections which usually cover a wide kinematic range, our lattice data are limited to low-energy scales of $1/z_0$ and finite range of $\lambda=zP^z$, so it is difficult for us to control the systematic uncertainties from model dependence like the global analyses. Nevertheless, we can still explore several models that are encoded with the power-law behavior $(1-x)^\beta$, and see how $\beta$ changes under threshold resummation. As we did previously in the NLO fit in Ref.~\cite{Gao:2020ito}, we use a simple {\it ansatz} for the pion valence PDF:
\begin{align} \label{eq:model}
 f_v^\pi(x;\alpha,\beta) &= {\cal N}x^\alpha\left(1-x\right)^\beta,
\end{align}
where $\mathcal{N}(\alpha,\beta)=\Gamma \left(2+\alpha+\beta \right)/(\Gamma (1+\alpha)\Gamma (1+\beta))$ is chosen such that $\int_0^1 dx f_v^\pi(x)=1$.

We construct moments from the models, which are determined by the parameters in the {\it ansatz}, and plug them into \eq{ratiotmc}. Then we fit the lattice data by minimizing
\begin{align}
&\chi^2 \equiv\\
& \sum_{P^z>P^z_0}^{P^z_{\rm max}}\sum_{z=z_{\rm min}}^{z_{\rm max}}\!\frac{\left[{\cal M}(z,P^z,P^z_0)\!-\!{\cal M}_{\rm model}(z,P^z,P^z_0;\alpha,\ldots)\right]^2}{\sigma^2_{\rm stat}(z,P^z,P^z_0)+\sigma_{\rm sys}^2(z,P^z,P^z_0)}\,,\nn
\end{align}
where $\sigma^2_{\rm stat}(z,P^z,P^z_0)$ are the statistic errors and $\sigma_{\rm sys}^2(z,P^z,P^z_0)$ are the systematic errors defined by scale variation,
\begin{align}
    \sigma_{\rm sys}(z) &\!=\!\frac{1}{2}\bigg{[}{\cal M}_{\rm model}(z)\Big|_{\alpha_s({\mu\over 2})} \!-\!  {\cal M}_{\rm model}(z)\Big|_{\alpha_s(2\mu)}\bigg{]}\,.
\end{align}
More details of a similar procedure can be found in Ref.~\cite{Gao:2020ito}.

\begin{figure}
\centering
	\includegraphics[width=0.95\linewidth]{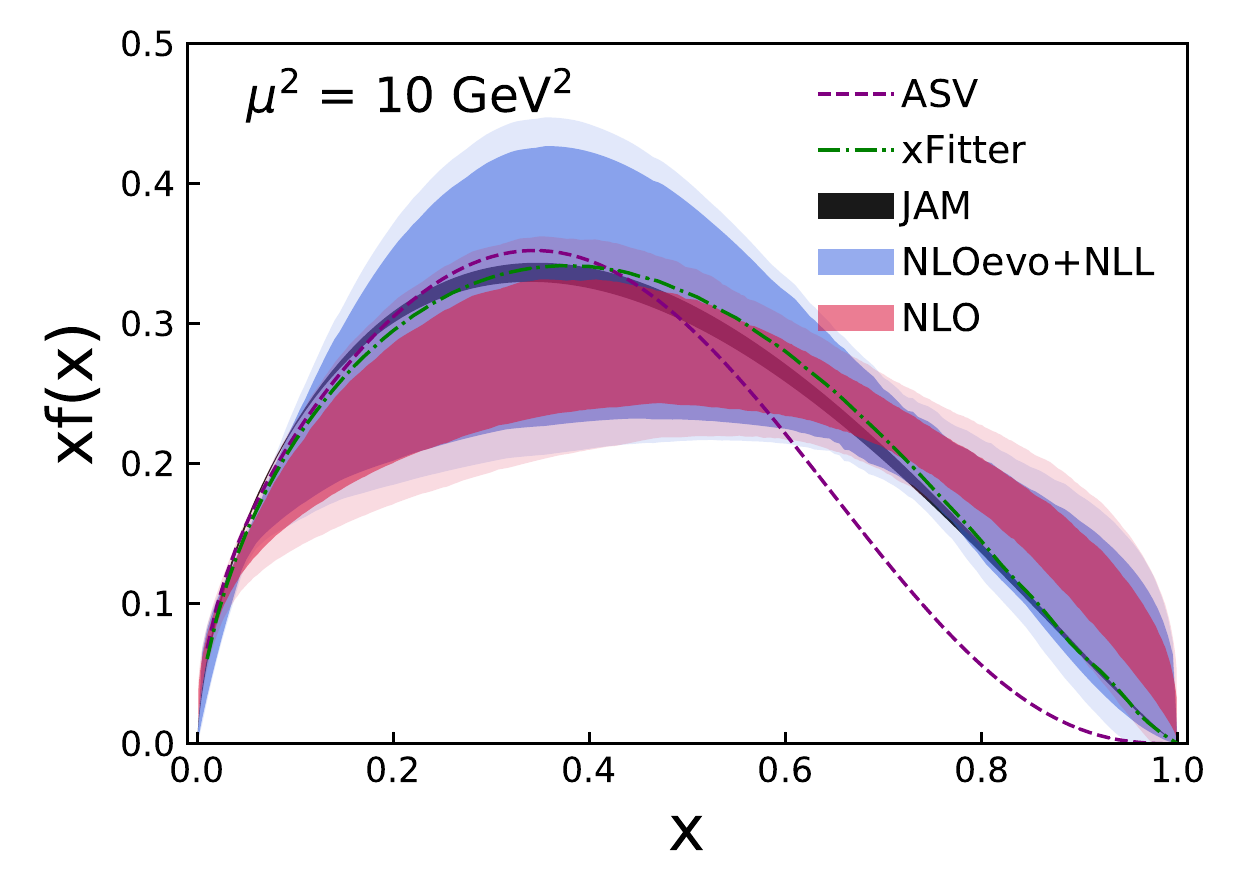}
	\caption{Valence PDF of pion from NLOevo+NLL fits of lattice results~\cite{Gao:2020ito} to \Eq{model}, with the darker and ligher bands being the statistic and systematic errors. Also shown are the same PDF from NLO fits~\cite{Gao:2020ito}, {\UrlFont xFitter}~\cite{Novikov:2020snp}, JAM~\cite{Barry:2018ort} and ASV~\cite{Aicher:2010cb}. \label{fig:PDFcmp}}
\end{figure}

Though the models are used to fit the lattice data, the corresponding parameters are mainly determined by the first few moments. Therefore, we truncate the series in \eq{ratiotmc} at $N = 16$ for the fit, which is large enough for the result to stabilize. Similar to the moment fit, we use $n_z^0$ = 1, 2, $z_{\rm min} = 2a, 3a$, and vary $z_{\rm max}\in [0.36, 0.48]$ fm to estimate the statistic and systematic errors. The results are shown in \tbl{conti}. As one can see, the large-$x$ exponent $\beta$ in this NLO+NLL analysis with DGLAP evolution (NLOevo+NLL) is around $1.0$ with about 50\% uncertainty. The central value of $\beta$ is increased from our previous fix-order NLO analysis~\cite{Gao:2020ito} by a small amount, which should be mainly driven by the DGLAP evolution, but they are still consistent within errors. This is due to the same reason that our current lattice data are only sensitive to the first few moments, which endures a small impact from threshold resummation. The situation can be improved if we manage to get more precise data and increase the pion momentum. One may notice that the results we get here have larger errors than our previous NLO fit~\cite{Gao:2020ito}, which is because we used more data with $z_{\rm max}$ up to 0.6 fm in that analysis. Finally, in \fig{PDFcmp} we show the comparison of pion valence PDF by {\UrlFont xFitter}~\cite{Novikov:2020snp}, JAM~\cite{Barry:2018ort}, ASV~\cite{Aicher:2010cb} as well as the fix-order estimate from the same lattice data~\cite{Gao:2020ito}.

\begin{table*}[!t]
\centering
\begin{tabular}{ l c c c c c c c r } 
\hline
  method & $a$ (fm) & $\langle x\rangle_v$& $\langle x^2\rangle_v$ & $\langle x^3\rangle_v$ & $\langle x^4\rangle_v$ & $\alpha$ & $\beta$ \\
\hline
  model &  0.06 &      & 0.1312(46)(32) &       & 0.0387(73)(47) &   &    \\
  independent &  0.04 &      & 0.1162(56)(19) &       & 0.0386(63)(31)  &   &      \\
  moments &  $a\to0$ &      & 0.1034(110)(32) &       & 0.0341(122)(47)  &      &  \\ [3mm]
  2-parameter &  0.06 & 0.282(18)(04) & 0.1352(56)(35) & 0.0801(55)(39) & 0.0533(63)(37) & -0.21(21)(06) & 0.97(39)(16) \\
  fit to &  0.04 & 0.247(21)(04) & 0.1189(60)(25) & 0.0720(57)(36) & 0.0492(65)(38) & -0.41(20)(06) & 0.74(44)(17) \\
  $\mathcal{N}x^\alpha(1-x)^\beta$&  $a\to0$ & 0.226(30)(06) & 0.1027(110)(26) & 0.0593(69)(28) & 0.0390(61)(28) & -0.40(24)(06) & 0.96(52)(15) \\
\hline
\end{tabular}
\caption{
Summary of results for valence PDF of pion at the scale $\mu=3.2$~GeV, obtained using NLO+NLL matching with DGLAP evolution (denoted as NLOevo+NLL) on the lattice QCD results of Ref.~\cite{Gao:2020ito}. We show results for two different lattice spacings $a$ as well as the continuum estimate denoted by $a\to 0$. For comparison, the 2-parameter fit results with NLO matching and without DGLAP evolution are $\alpha$ = -0.55(15)(08) and $\beta$ = 0.66(34)(22) in \refcite{Gao:2020ito}.
}
\label{tbl:conti}
\end{table*}

\section{Conclusion}
\label{sec:sum}

In this work, we studied the threshold resummation in the qPDF, the pPDF and the corresponding spatial correlator, which have been used for the lattice calculation of PDFs in the LaMET and pPDF approaches. We find that the threshold logarithms in them originate from the interplay between the collinear and soft divergences in a 3D momentum distribution, and they are sensitive to long-range correlations of the spatial correlator in coordinate space. Moreover, the 3D momentum distribution, when Fourier transformed from $k_\perp$ to $b_\perp$ space, reduces to the qPDF and the pPDF
in the $b_\perp\to0$ and infinite momentum limits, respectively, which offers us new insights on the relation between the latter two quantities.

Motivated by the standard procedure, we propose to resum these threshold logarithms in the OPE of the spatial correlator at NLL accuracy, and confirm our results by comparing them with the NNLO Wilson coefficients. Unlike DIS and DY, the resummed Wilson coefficients in this case match the factorization scale to the UV fixed point in the threshold limit, and therefore are free from the influence of Landau poles. In addition, we included DGLAP evolution for applications to more realistic lattice calculations where the range of available $z$ is not small. We showed that, apart from the DGLAP evolution, our proposed resummation of threshold logarithms leads to the suppression of PDFs at large $x$, similar to the DIS and DY cases as observed in the ASV analysis~\cite{Aicher:2010cb}. 

We then reanalyzed the pion valence PDF using recent lattice data from~\cite{Gao:2020ito} by applying our proposed resummation formula at NLO+NLL accuracy with LL DGLAP evolution. We found that threshold resummation makes little impact on the second and fourth Mellin moments, as well as on the large-$x$ exponent of the PDF from model fits, compared to the previous fixed-order analysis~\cite{Gao:2020ito} with NLO matching. This is because the currently available lattice data are sensitive only to the lowest moments or moderate values of $x$, and they are not precise enough. On the other hand, the effects of DGLAP evolution in the NLO matching constitutes an important systematic uncertainty that should be included in the analysis.
Finally, with increased pion momentum and data precision in the future, we expect threshold resummation to become a necessary systematic correction in the calculation of higher moments and the large-$x$ PDF.

\begin{acknowledgments}

We thank Felix Ringer for early communications on threshold logarithms for the qPDF, and Zheng-Yang Li and Yan-Qing Ma for providing the two-loop matching coefficient for the spatial correlator. We would like to thank Patrick Barry and George Sterman for their helpful discussions. We also thank Andreas Sch\"{a}fer, George Sterman, Werner Vogelsang, and Feng Yuan for their feedback on the manuscript. KL thanks Margaret Mariano for the hospitality while this work was completed.

This material is based upon work supported by: (i) The U.S. Department of Energy, Office of Science, Office of Nuclear Physics through the Contract No.~DE-SC0012704, No.~DE-AC02-06CH11357 and No.~DE-SC007976; (ii) The U.S. Department of Energy, Office of Science, Office of Nuclear Physics and Office of Advanced Scientific Computing Research within the framework of Scientific Discovery through Advance Computing (ScIDAC) award Computing the Properties of Matter with Leadership Computing Resources; (iii) The U.S. Department of Energy, Office of Science, Office of Nuclear Physics, within the framework of the TMD Topical Collaboration. (iv) X.G. is partially supported by the NSFC Grant Number 11890712. (vi) K.L. is partially supported by the Department of Energy, Office of Nuclear Physics. (i) This research used awards of computer time provided by the INCITE and ALCC programs at Oak Ridge Leadership Computing Facility, a DOE Office of Science User Facility operated under Contract No. DE-AC05-00OR22725. (ix) Computations for this work were carried out in part on facilities of the USQCD Collaboration, which are funded by the Office of Science of the U.S. Department of Energy.

\end{acknowledgments}

\appendix

\section{Leading collinear and soft divergence in the momentum space}
\label{app:1loopsail}

\begin{figure}[!th]
 \centering
    \includegraphics[width=0.45\linewidth]{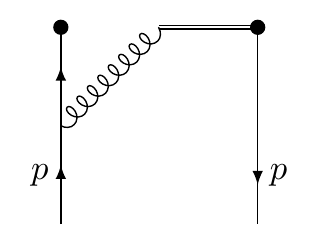}
    \includegraphics[width=0.45\linewidth]{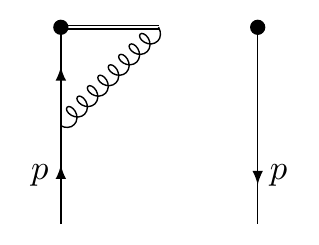}
  \caption{One-loop Feynman diagrams that contribute to the leading collinear and soft divergence of the 3D momentum distribution. The conjugate diagrams are implied as well.}
  \label{fig:sail}
\end{figure}

In this section, we calculate the 3D momentum distribution at one-loop. The leading collinear and soft divergence comes from the Feynman diagrams in \fig{sail}.

The Feynman rule for these diagrams is
\begin{align}
    \tilde q(x,\vec{k}_\perp,p^z)&\!=\! 2g^2 C_F \int {d^{d-1}\vec{b}\over (2\pi)^{d-1}} e^{i\vec{k}\cdot \vec{b}}\int {dl^0 d^{d-1}\vec{l}\over (2\pi)^d}\nn\\
    &\quad \times { l^0 z \!+\! \vec{l}\cdot \vec{b} \over l^2 (p-l)^2 }  \int_0^1 ds \ e^{-i(\vec{p}-\vec{l})\cdot \vec{b}s} e^{-i\vec{l}\cdot \vec{b}}\,,
\end{align}
where $s$ is the path parameter for the Wilson line, and the Fourier transform is in $(d-1)$-dimensional space.

By using the identity,
\begin{align}
    l^0z+\vec{l}\cdot \vec{b}=(l^0+p^z)z-(\vec{p}-\vec{l})\cdot \vec{b}\,,
\end{align}
we can drop the second term on the right-hand side, for it does not suffer from divergences as $x\to 1$. Keeping the first term, we can integrate over $k^0$ and obtain $\tilde q_{\rm cs}(x,\vec{k}_\perp,p^z)$ whose real part is
\begin{align}\label{eq:qreal}
    &\tilde q_{\rm cs}(x,\vec{k}_\perp,p^z)\\
    &= {g^2 C_F\over 2(2\pi)^{d-1}} \int_0^1 ds\ {(1-s)^{2-d}\over \vec{k}_\perp^2}\nn\\
    &\times \left[\frac{k_t^2 (1\!+\!x\!-\!2 s)\!+\!(x-s)^3}{\left( k_t^2+(s-x)^2\right)^{3/2}}\!-\!\frac{k_t^2 (\!1\!+\!x\!-\!2 s)\!+\!(x-1)^3}{\left( k_t^2+(x-1)^2\right)^{3/2}}\right]\,,\nn
\end{align}
where the $(1-s)^{2-d}$ term originates from the $(d-1)$-dimensional integration over $\vec{b}$, and we keep $\epsilon$ finite throughout the calculation.

Integrating over $k_\perp$ while keeping $p^z$ finite, we obtain the contribution to the qPDF,
\begin{align}\label{eq:qreal1}
    &\tilde q_{\rm cs}(x,p^z)= \int d^{d-2}k_\perp \ \tilde q_{\rm cs}(x,\vec{k}_\perp,p^z)\\
    &= -{g^2 C_F\over 16\pi^2}{\Gamma(\eps+{1\over2})\over \sqrt{\pi}} {1\over \eps(1-2\eps)}\nn\\
    &\quad \times {(1-x)^{-2\eps}(1+x-4\eps) + x^{-2\eps}(1+x-2\eps)\over 1-x}\,.\nn
\end{align}

In the limit of $k_t^2\to0$, \eq{qreal} reduces to
\begin{align}\label{eq:qreal2}
    &\tilde q_{\rm cs}(x,\vec{k}_\perp,p^z)\nn\\
    &=  {g^2 C_F\over 2(2\pi)^{d-1}}\left[ \int_0^1 ds\ {(1-s)^{2-d}\over \vec{k}_\perp^2}\left(\frac{ x-s }{|x-s|}+\frac{1-x}{|1-x|}\right)\right.\nn\\
    &\left.\qquad + \int_0^1 ds\ (1-s)^{2-d} \delta(x-s) {2(1-s)\over \vec{k}_\perp^2}\right]\nn\\
    &={g^2 C_F\over (2\pi)^{d-1}} {1\over \vec{k}_\perp^2} \left[-1 + (1-\eps){2(1-x)^{2\eps}\over (1-x)}\right]{1\over 1-2\eps}\nn\\
    &\qquad\times \theta(x)\theta(1-x)\,,
\end{align}
where the $\delta(x-s)$ term in the first ``='' comes from the singularity when $k_t^2\to0$ and $x\to s$ in the first term in the square bracket in \eq{qreal}, and its coefficient is obtained by integrating over $x$ from $-\infty$ to $\infty$. The limit $k_t^2\to0$ restricts $x$ to be in the physical region $[0,1]$, and then we obtain \eq{leaddiv1} in the soft limit $x\to 1$.

\bibliography{lattPDF}

\end{document}